\documentclass[fleqn,usenatbib]{mnras}

\usepackage{newtxtext,newtxmath}

\usepackage[T1]{fontenc}

\DeclareRobustCommand{\VAN}[3]{#2}
\let\VANthebibliography\thebibliography
\def\thebibliography{\DeclareRobustCommand{\VAN}[3]{##3}\VANthebibliography}

\usepackage{newtxtext,newtxmath}
\usepackage{fontawesome5}
\usepackage{mathtools}
\usepackage{float}
\usepackage{graphicx}
\usepackage{xtab,booktabs}
\usepackage{longtable}
\usepackage{lipsum}
\usepackage{natbib}

\newcommand{\thisgrb}{GRB~201221D }
\newcommand{\tninty}{{$T_{\rm 90}$} }
\newcommand{\swift}{{\it Swift} }
\newcommand{\swiftT}{{T$_{\rm 0}$} }
\newcommand{\fermi}{{\it Fermi} }
\newcommand{\gbm}{{\it Fermi}/GBM }
\newcommand{\bat}{{\it Swift}/BAT }

\newcommand{\kw}{Konus-{\em Wind} }

\newcommand{\fermiT}{{T$_{0}$} }
\newcommand{\keV}{{\rm keV} }

\newcommand{\Ep}{$E_{\rm p}$ }

\newcommand{\sw}[1]{\texttt{#1}}
\newcommand\T{\rule{0pt}{4ex}}  

\title[Short GRB~201221D]{Multi-wavelength analysis of short \thisgrb and its comparison with other high \& low redshift short GRBs}

\author[Dimple et al.]{ 
Dimple$^{1,2}$\thanks{E-mail: dimplepanchal96@gmail.com, dimple@aries.res.in},
K. Misra$^1$, D. A. Kann$^3$, K. G. Arun$^4$, A. Ghosh$^{1,5}$, R. Gupta$^{1,2}$,
L. Resmi$^6$,
J. F. Ag\"u\'i Fern\'andez$^3$, 
\newauthor 
C. C. Th\"one$^3$,
A. de Ugarte Postigo$^{7}$, S. B. Pandey$^1$, L. Yadav$^2$
\\
\\
$^1$ Aryabhatta Research Institute of Observational Sciences (ARIES), Manora Peak, Nainital-263002, India. \\
$^2$ Department of Physics, Deen Dayal Upadhyaya Gorakhpur University, Gorakhpur-273009, India. \\
$^3$ Instituto de Astrof\'isica de Andaluc\'ia, Glorieta de la Astronom\'ia s/n, 18008 Granada, Spain.\\
$^4$ Chennai Mathematical Institute, Siruseri, 603103 Tamilnadu, India.\\
$^5$ School of Studies in Physics and Astrophysics, Pandit Ravishankar Shukla University, Chattisgarh, 492 010, India.\\
$^6$ Indian Institute of Space Science and Technology, Trivandrum 695 547, India.\\
$^7$ Artemis, Observatoire de la Côte $D^,$Azur, Universite\'  Côte $D^,$Azur, CNRS, Nice, 06300, France\\}

\date{Accepted XXX. Received YYY; in original form ZZZ}

\pubyear{2021}

\begin{document}
\label{firstpage}
\pagerange{\pageref{firstpage}--\pageref{lastpage}}
\maketitle

\begin{abstract}
We present a detailed analysis of short \thisgrb lying at redshift $z= 1.045$. We analyse the high-energy data of the burst and compare it with the sample of short gamma-ray bursts (SGRBs). The prompt emission characteristics are typical of those seen in the case of other SGRBs except for the peak energy ($E_{\rm p}$), which lies at the softer end (generally observed in the case of long bursts). We estimate the host galaxy properties by utilising the \sw{Python}-based software \sw{Prospector} to fit the spectral energy distribution of the host. The burst lies at a high redshift relative to the SGRB sample with a median redshift of $z=0.47$. We compare the burst characteristics with other SGRBs with known redshifts along with GRB~200826A (SGRB originated from a collapsar). A careful examination of the characteristics of SGRBs at different redshifts reveals that some of the SGRBs lying at high redshifts have properties similar to long GRBs indicating they might have originated from collapsars. Further study of these GRBs can help to explore the broad picture of progenitor systems of SGRBs.

\end{abstract}



\section{Introduction}
\label{sec:intro}

The bi-modality in duration distribution of Gamma-Ray Bursts (GRBs) revealed two broad populations identified as short and long GRBs (based on\tninty\footnote{\tninty is the duration over which a particular instrument observes 5\% to 95\% of the total counts.} duration with separation boundary at 2 sec, \citealt{Mazets1981ApSS, Kouveliotou1993}). The two GRB populations are likely originating from two distinct progenitor systems, with different redshift distribution and located in diverse host galaxy environments \citep{nakar2007, Berger2014, Levan2016}. The association of long GRBs with broad-lined supernovae of Type Ic and their occurrence in star-forming galaxies confirmed their association with collapsars \citep{Woosley_1993, MacFadyen1999, Hjorth2003, Woosley2006, Li_2016}. On the other hand, a mix of young and old stellar population of host galaxies of SGRBs and the lack of associated supernova suggests that at least a fraction of SGRBs originate from compact object mergers \citep{2009ApJ...690..231B, Fong2013, Beniamini2016a}. The discovery of gravitational wave signal GW170817 and its association with SGRB~170817A confirmed this hypothesis \citep{Abbott_2017, Goldstein2017, valenti_2017}. 

However, some of the long GRBs (like GRBs~060614 and 060505) have no evidence of supernova association despite long follow-up \citep{Della_2006, Fynbo_2006, Gal_2006}. Similarly, signatures of collapsars are seen in some of the SGRBs (for example, SGRBs~090426 and 200826A,  \citealt{Antonelli_2009,Thoene2011MNRAS,NicuesaGuelbenzu2011AA,NicuesaGuelbenzu2012AA,2021arXiv210505067A, zhang_2021, 2021arXiv210503829R}). The absence of supernova signatures in long GRBs and the occurrence of SGRBs from collapsars challenge our current understanding of GRB population and their progenitor systems. Several attempts have been made in the past to devise new classification schemes based on different criteria other than \tninty. \citet{Zhang_2006} divided the GRBs into Type I (compact star origin) and Type II (massive star origin) classes. \citet{Bromberg2013} classified the GRBs as collapsars and non-collapsars based on the non-collapsar probability. Later, \citet{M2019} used the $E_{\rm \gamma, iso}$ - $E_{\rm p,i}$ correlation to divide the GRBs in two classes. These works have allowed to develop a classification scheme beyond the traditional \tninty distribution.

The distance measurement of the bursts can also provide essential information about their intrinsic energy budgets, the progenitor age distribution, and its relation to star-formation \citep{Guetta2005, Berger2007, Ghirlanda2009AA, Avanzo2015}. Therefore, the redshift distribution of GRBs serves as a clue to the progenitor systems. SGRBs are generally found at low redshifts (with a median redshift $\overline{z}=0.47$) compared to long GRBs (with a median redshift $\overline{z}=1.68$; see section \ref{comparison} for details). The redshift distribution of SGRBs can be explained through their formation channel. The time taken by compact objects to merge (through energy/angular momentum loss by GW radiation) is quite long \citep{ Belczynski2006, Beniamini2016c}. Therefore, if SGRBs originate from compact object mergers, they are more likely to lie at lower redshifts. However, a fraction of SGRBs are found to be located at high redshifts \citep{2006ApJ...648L..83D,Berger2007}.

It has also been observed that SGRBs at $z>1$ have a high probability of being collapsars \citep{Bromberg2013}. It is also interesting to note that both the SGRBs~200826A \citep[$z=0.7481$;][]{2021arXiv210503829R} and 090426 \citep[$z=2.609$;][]{Antonelli_2009}, which have been found to originate from the death of massive stars, lie at the higher end of the redshift distribution of SGRBs. \thisgrb is  located at the higher end of the GRB redshift distribution \citep[$z=1.045$,][]{fernandez2021grb}, which gives rise to the question if the burst originates from a collapsar or a merger? In general, it is vital to investigate if the SGRBs lying at high redshifts have progenitor systems similar to the SGRBs lying at low redshifts? To address this question and the progenitor conundrum, we compare the properties of SGRBs in the context of the available redshift information.

The paper presents a detailed analysis of \thisgrb and its comparison with other low and high redshift SGRBs. The data reduction procedure and analysis are described in \S\ref{Data acquisition and analysis}. The results obtained are discussed in \S\ref{results}, including the properties of the host galaxy. In addition, we compare the SGRBs with known redshift to identify the similarities and differences between high and low redshift SGRB samples in \S\ref{comparison}. A brief summary of this work is presented in \S\ref{conclusions}. We quote all the uncertainties at $1\sigma$ throughout this paper (unless otherwise mentioned). We used the Hubble parameter $\rm H_{0}$ = 70 km $\rm sec^{-1}$ $\rm Mpc^{-1}$, and the density parameters $\rm \Omega_{\Lambda}= 0.73$, and $\rm \Omega_m= 0.27$ in this paper. The measured redshift of $z=1.045$ corresponds to a luminosity distance of 7109~Mpc.

\section{Data acquisition and analysis}
\label{Data acquisition and analysis}
\swift triggered on \thisgrb on December 21 2020, with the burst having a duration of 0.3 sec \citep{2020GCN.29112....1P}. The \fermi and \kw missions also detected the burst \citep{2020GCN.29140....1H,2020GCN.29130....1F}. Later, various ground-based telescopes started observations of the burst location to search its optical counterpart. Spectroscopic observations of the optical counterpart of \thisgrb with the Gran Telescopio Canarias (GTC) provided the measurement of the redshift of $z=1.045$ \citep{2020GCN.29132....1D}. We also observed the burst location with the 3.6m Devasthal Optical Telescope (DOT) and detected an extended source at the location of the burst \citep{2020GCN.29148....1P}. 
 
This section describes the data acquisition and analysis, using the data from different space and ground-based instruments, in the prompt emission and afterglow phase.

\subsection{{\it Swift}/BAT}
\thisgrb triggered the Burst Alert Telescope \citep[BAT;][]{Bat_barthelmy2005} on-board the \textit{Neil Gehrels Swift Observatory} (\swift hereafter) on December 21, 2020 at 23:06:34 UT. The best localization of the source was found to be at RA: 11h 24m 12s and Dec: +42d 08m 39s (J2000) with an uncertainty radius of 3\arcmin{} \citep{2020GCN.29112....1P}.  

\begin{figure}
\centering
\includegraphics[width=\columnwidth]{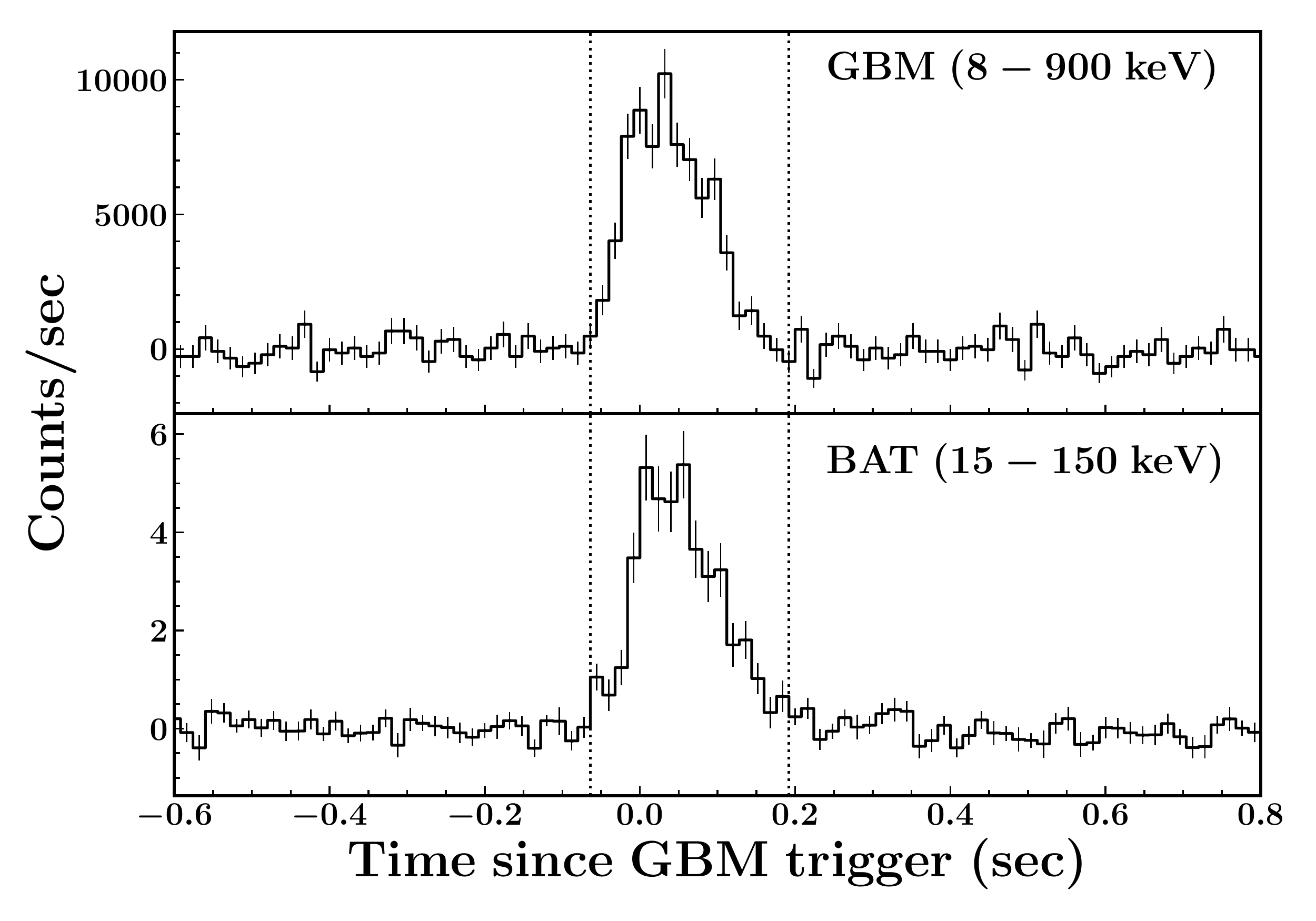}
\caption{Light curves of \thisgrb with a resolution of 64 ms in the energy ranges (8-900) \keV and (15-150) \keV using \gbm and \bat data, respectively. The dotted lines show the start and end times of the transient.}
\label{batgbmlc}
\end{figure}

To extract the temporal and spectral features from the \bat data, we obtained the raw data from the \swift Archive Download Portal supported by the UK \swift Science Data Centre\footnote{\url{https://www.swift.ac.uk/swift_portal/}}. We utilized \sw{HEASOFT} version-6.25 with the latest \swift calibration data files\footnote{\url{https://heasarc.gsfc.nasa.gov/FTP/caldb/}} to reduce this data. The three primary tools, namely \sw{batbinevt}, \sw{bathotpix} and \sw{batmaskwtevt} were used to create the Detector Plane Image (DPI), to detect the hot pixels, and for mask-weighting, respectively. The mask-weighted light curve in the $15-150$ keV energy range is extracted using \sw{batbinevt}. The bottom panel of Fig. \ref{batgbmlc} shows the \bat light curve. The light curve consists of a single-peaked structure with a duration $T_{90}=0.16\pm0.04$ sec \citep{2020GCN.29112....1P, 2020GCN.29139....1K}.

Furthermore, we obtained the time-averaged spectrum in a time interval starting from \swiftT-0.064 sec to \swiftT+0.192 sec following the method specified in the \bat software guide\footnote{\url{https://swift.gsfc.nasa.gov/analysis/bat_swguide_v6_3.pdf} }. The \sw{pha} and response files obtained are used for joint spectral analysis along with \fermi data (see section \ref{joint_spectrum}).
 
\subsection{{\it Fermi}/GBM}
The Gamma-Ray Burst Monitor (GBM, \citealt{Meegan2009}) onboard the \fermi spacecraft triggered and located \thisgrb at 23:06:34.33 UT. Initially, the flight software classified the trigger as a particle event. Later, it was confirmed to be an SGRB with a \tninty duration of about 0.14 sec (50-300 keV). The burst location provided by \fermi was consistent with the \bat position \citep{2020GCN.29140....1H}. We used the time-tagged event (TTE) data of GBM obtained from the GBM trigger data archive\footnote{\url{https://heasarc.gsfc.nasa.gov/FTP/fermi/data/gbm/triggers/}} for spectral and temporal analysis of the burst in the high-energy regime. We chose the detectors with low observing angles and high count rates. Three sodium iodide (NaI) detectors: n7, n8, and nb were selected by visually inspecting the count-rate light curves and source observing angles (n7 -- $\rm 43^\circ$, n8 -- $\rm 5^\circ$, nb -- $\rm 57^\circ$). One of the bismuth germanate detectors (BGO1 -- $\rm 61^\circ$), closer to the direction of burst) was also included in our analysis. 

\begin{figure}
\includegraphics[width=\columnwidth]{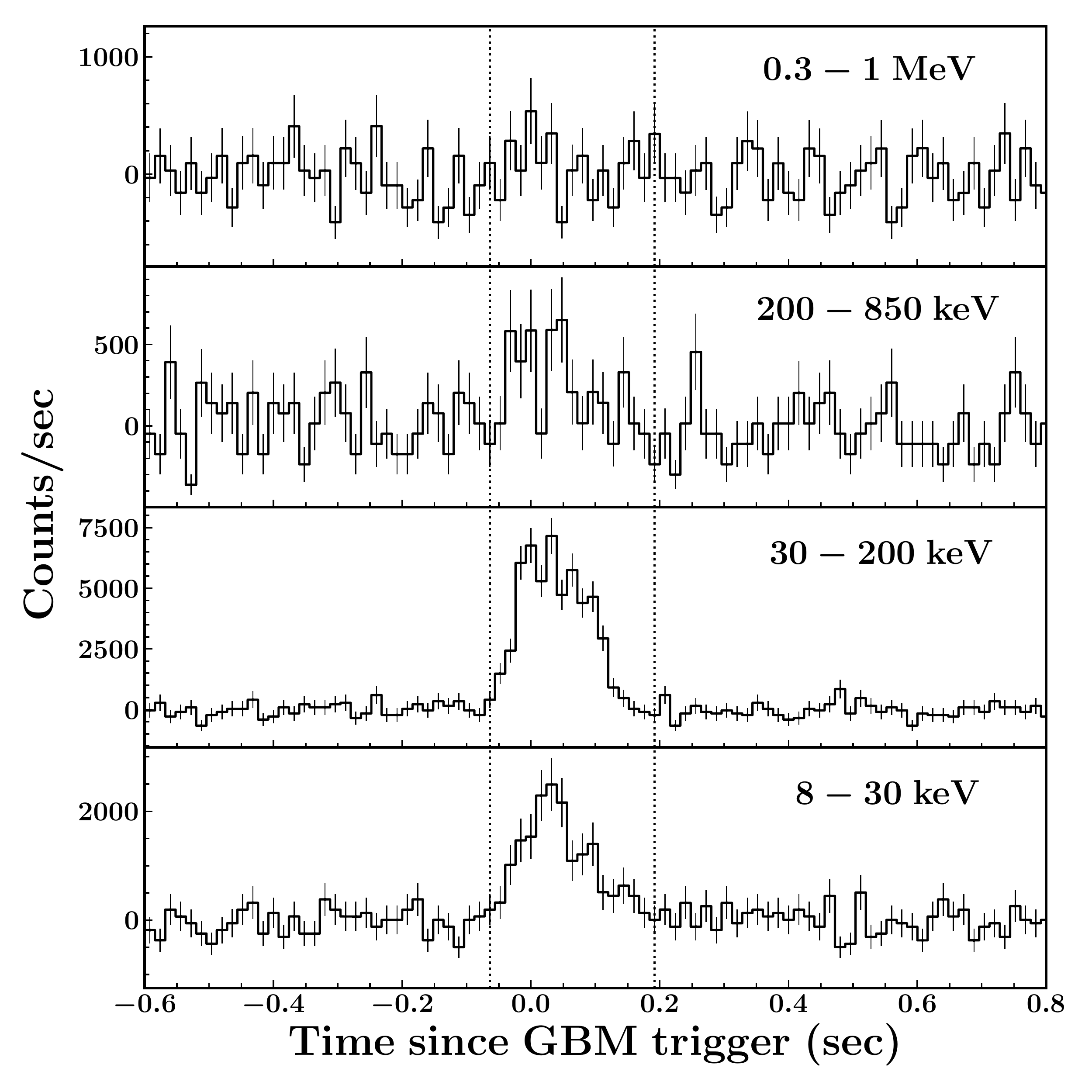}
\caption{{Prompt emission light curves of \thisgrb in different energy channels of \gbm with a time resolution of 64 ms. The burst duration in higher energy channels is shorter than that in lower energy channels.} }
\label{promptlc}
\end{figure}

We used \sw{RMFIT}\footnote{\url{https://fermi.gsfc.nasa.gov/ssc/data/analysis/rmfit/}} (version 4.3.2) to visualize the light curves from the TTE files. From these light curves, we carefully selected the source and background. We fitted the background with various polynomial functions. The best-fitted background was subtracted from the source to produce light curves in different energy bins. The background-subtracted multi-channel prompt emission $\gamma$-ray/hard X-ray light curves are shown in Fig. \ref{promptlc}. 

\par
For spectral analysis, the background-fitted time-averaged spectrum for the time bin between \fermiT-0.064 to \fermiT+0.192 sec was obtained using the \sw{GTBurst} software from the \emph{Fermi Science Tools}. 
The \sw{pha} files obtained are used for joint spectral analysis along with \swift data (see section \ref{joint_spectrum}).

\subsection{ {Joint \swift and \fermi spectral analysis}}
\label{joint_spectrum}
To investigate the emission mechanism of GRB~201221D, we performed a joint spectral analysis of \gbm and \bat data using \sw{threeML} \citep[\sw{3ML},\footnote{\url{https://threeml.readthedocs.io/en/latest/}}][]{Vianello2015} version 2.3.1. Joint spectral analysis was done utilizing the \gbm spectrum over the energy range of $8-900$ keV (for NaI) and $200-30000$ keV (BGO) and the \bat data with energy range $15-150$ keV. We removed the $33-37$ keV energy channels to ignore the K-edge (33.17 keV) of the Na line from the spectral analysis of NaI data. We tried to fit the spectrum with a power-law function having an exponential cutoff (\sw{CPL} model),  \sw{Band} function and \sw{Black Body} along with \sw{Band} function. Based on the Bayesian Information Criteria \citep[BIC;][]{Kass:1995}, Akaike Information Criteria (AIC), and Log(likelihood) for each model, we found that the spectrum is best described with a \sw{CPL} with power-law index of $-0.20\pm0.16$ and cutoff energy $E_{c}=51.14_{-6.7}^{+7.2}$ \keV, which is re-parameterized to $E_p=110.4_{-13}^{+14}$ \keV with a fluence of $(1.02\pm0.1)\times 10^{-6}$ erg cm$^{-2}$, consistent with the values reported by \citet{2020GCN.29140....1H}. 

\begin{table}
\caption{The best-fitting models and the spectral parameters obtained from time-resolved spectroscopy of GRB~201221D.}
\begin{tabular}{ c c c c l }

\hline
\bf {Time interval} \ \ & \bf{Model} &  \ {\boldmath${\alpha}$} & \
{$\bf {\it E}_{p}$} & \ \ \bf {Flux}\\
\textbf{(sec)} & & & \textbf{(\keV)} & \textbf{ ($ \bf 10^{-6}\ erg/s/cm^{2}$)}\\
\hline
\T
$-0.044$ -- $-0.005$ & CPL & $-0.34_{-0.54}^{+0.51}$ &  $47_{-15}^{+22}$  &  $2.6_{-2.1}^{+22}$ \\
\T
$-0.005$ --  $0.112$ & CPL &  $-0.37_{-0.19}^{+0.20}$ &  $45_{-5}^{+6}$ &   $3.8_{-2.4}^{+7.0}$\\
\T
$0.112$ --  $0.191$ & CPL & $-1.09_{-0.66}^{+0.67}$  & $18_{-5}^{+7}$ & $0.34_{-0.32}^{+12}$\\
\hline
\end{tabular}
\label{tab:TRS}
\end{table}

\subsubsection{ {Time-resolved spectroscopy}}
For time-resolved spectral analysis, we created the time bins from background-subtracted \gbm light curves by applying the bayesian blocks \citep{Scargle_2013} to the main emission interval (\fermiT-0.064 to \fermiT+0.192 s). We used the NaI-8 detector with the maximum count rate and obtained four Bayesian bins. However, we could use only three bins for spectral analysis as the first bin did not have sufficient counts to be modelled. We created the spectra for three bins and fitted them with various models (\sw{Band}, \sw{Black Body}, and \sw{CPL}). We found that all of these spectra are well described with the \sw{CPL} function. The best-fit model and the spectral parameters obtained from the time-resolved spectroscopy for \thisgrb are listed in Table \ref{tab:TRS}. 

\begin{figure}
\includegraphics[width=\columnwidth]{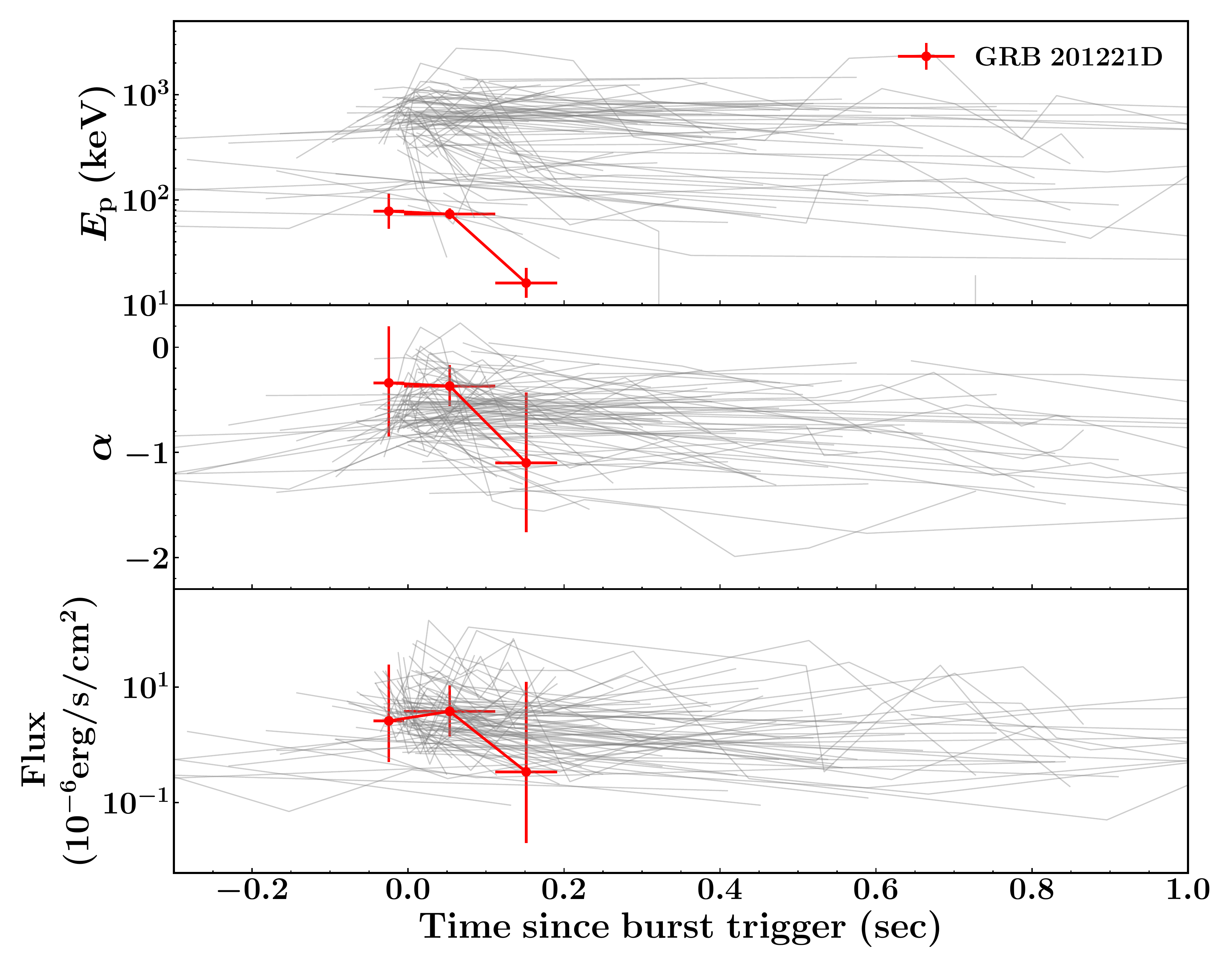}
\caption{Evolution of spectral parameters for \thisgrb and its comparison with the sample of SGRBs taken from \citet{burgess2017bayesian}. The photon index and flux values are typically comparable to the SGRB sample. However, the peak energy value of the burst is quite low in the last bin compared to the sample of SGRBs. All these parameters follow a hard-to-soft evolution.}
\label{TRS}
\end{figure}

The evolution of spectral parameters is shown in Fig. \ref{TRS}. All the parameters (flux, $\alpha$ and $E_{\rm p}$) are seen to follow the same evolution pattern. The figure also shows a comparison of the evolution of parameters with an SGRB sample presented in \citet{burgess2017bayesian}. The values of different parameters in the case of \thisgrb are typical, following a hard-to-soft evolution, as compared to the sample of SGRBs except for the cut-off energy, which lies at the lower end of the distribution.

\begin{table*}
\centering
\caption{AB magnitudes of the afterglow/host of \thisgrb. Magnitudes are not corrected for Galactic extinction.}
\begin{tabular}{c  c  c  c  l  l} 

 \hline
{\bf $\Delta$ t (days)}  &  \bf Filter  &\bf   Magnitude (AB) & \bf Telescope & \bf Reference \\
\hline
0.069  & $r^\prime$ & $23.10\pm0.30$ & NOT & \cite{2020GCN.29117....1M}\\
0.115  & $r^\prime$ & $23.95\pm0.20$ & GTC & \cite{fernandez2021grb}\\
0.400  & $J$ & $21.8\pm0.20$  & MMT & \cite{2020GCN.29142....1R}\\
0.421  & $r^\prime$ & $\sim 23.90$ & LMI & \cite{2020GCN.29128....1D}\\
0.421  & $i^\prime$ & $\sim 23.70$ & LMI & \cite{2020GCN.29128....1D}\\
0.997  & $r^\prime$ & $ 23.62\pm0.30$  &  DOT & This Work \\ 
13.879  & $J$ & $22.40\pm0.17^h$ & LBT & \cite{fernandez2021grb}\\
13.895  & $K_{s}$ & $22.15\pm0.20^h$ & LBT & \cite{fernandez2021grb}\\
19.349  & $g^\prime$ & $23.80\pm0.12^h$ & LBT & \cite{fernandez2021grb}\\
19.349  & $r^\prime$ & $23.83\pm0.15^h$ & LBT & \cite{fernandez2021grb}\\
19.349  & $i^\prime$ & $23.44\pm0.18^h$ & LBT & \cite{fernandez2021grb}\\
19.349  & $z^\prime$ & $23.11\pm0.25^h$ & LBT & \cite{fernandez2021grb}\\
-- & y & $22.6\pm0.20^h$ & Pan-STARRS & \cite{2020GCN.29133....1K} \\
165.75  & $R_C$ & $  > 23.20^h$  &  DOT & This Work \\ 
175.66  & $R_C$ & $  > 22.90^h$  &  HCT & This Work \\ 
\hline
$^h$ --  Host magnitudes\\
\end{tabular}

\label{tab:optical_data}  
\end{table*}
\subsection{{\it Swift}/XRT}
The X-ray telescope \citep[XRT;][]{Burrows2005} on-board \swift started observing the field at 23:08:01.7 UT, 87.4 sec after the BAT trigger.  A new, faint, uncatalogued X-ray source was detected at RA: 11h 24m 14.19s, Dec: +42d 08m 35.5s (J2000) with an uncertainty of 5\farcs7 (radius, 90\% containment). Due to the faintness of the source, XRT observed it only in Photon Counting (PC) mode. The X-ray afterglow light curve, available at the \swift online repository\footnote{\url{https://www.swift.ac.uk/}} provided by the University of Leicester \citep{eva07, eva09}, consists of only one data point (with a large error in time) followed by an upper limit. Further investigation of the X-ray afterglow could not be performed. However, in  \S\ref{comparison} we compare the X-ray light curves of SGRBs, including GRB~201221D.

\subsection{Optical}

\begin{figure}
\centering
\includegraphics[width=\columnwidth]{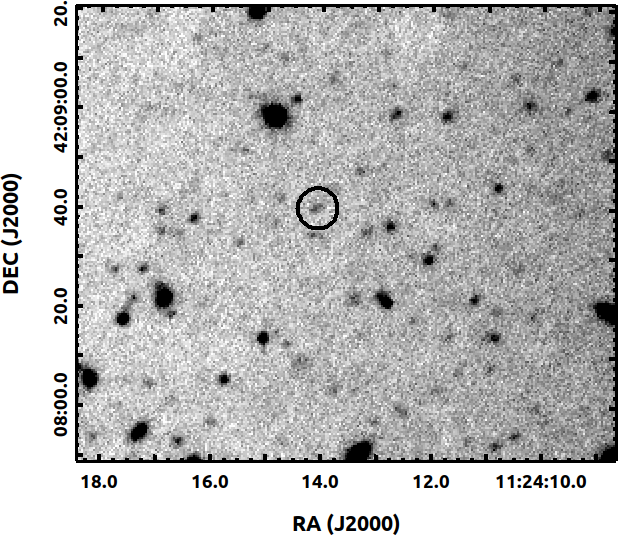}
\caption{Optical image of \thisgrb taken $\sim1$ day after the burst using ADFOSC mounted on the 3.6m DOT. An extended source can be clearly seen at the location of the burst.}
\label{FC}
\end{figure}

The optical afterglow emission of \thisgrb was discovered using the Nordic Optical Telescope (NOT) at $\sim1.67$ hr after the burst with $r^\prime=23.1\pm0.3$ mag \citep{2020GCN.29117....1M}. Spectroscopic observations with the GTC/OSIRIS at $\sim$2.76 hr after the burst showed evidence of absorption lines \citep{fernandez2021grb}, yielding a redshift $z=1.045$. This is only the third spectrum of a SGRB afterglow \cite[after GRB~130603B and GRB~160410A;][]{Cucchiara_2013,deUgarte_2014AA,fernandez2021grb} which displayed absorption-line features. The $r^\prime$-band acquisition image from GTC/OSIRIS detected the afterglow with a magnitude of $23.95\pm0.20$ mag \citep{fernandez2021grb}. A source was also identified in the observations with the Large Monolithic Imager (LMI) on the 4.3m Lowell Discovery Telescope in $r^\prime$ and $i^\prime$ bands at $\sim10.11$ hr \citep{2020GCN.29128....1D}. 
Further multi-band observations of the host galaxy were also performed with the Multiple Mirror Telescope (MMT) and Large Binocular Telescope (LBT) \citep{Rastinejad2021, 2021GCN.29311....1R, fernandez2021grb}. The optical/NIR magnitudes of the afterglow/host available in the literature are listed in Table \ref{tab:optical_data}.

\begin{figure}
\includegraphics[width=\columnwidth]{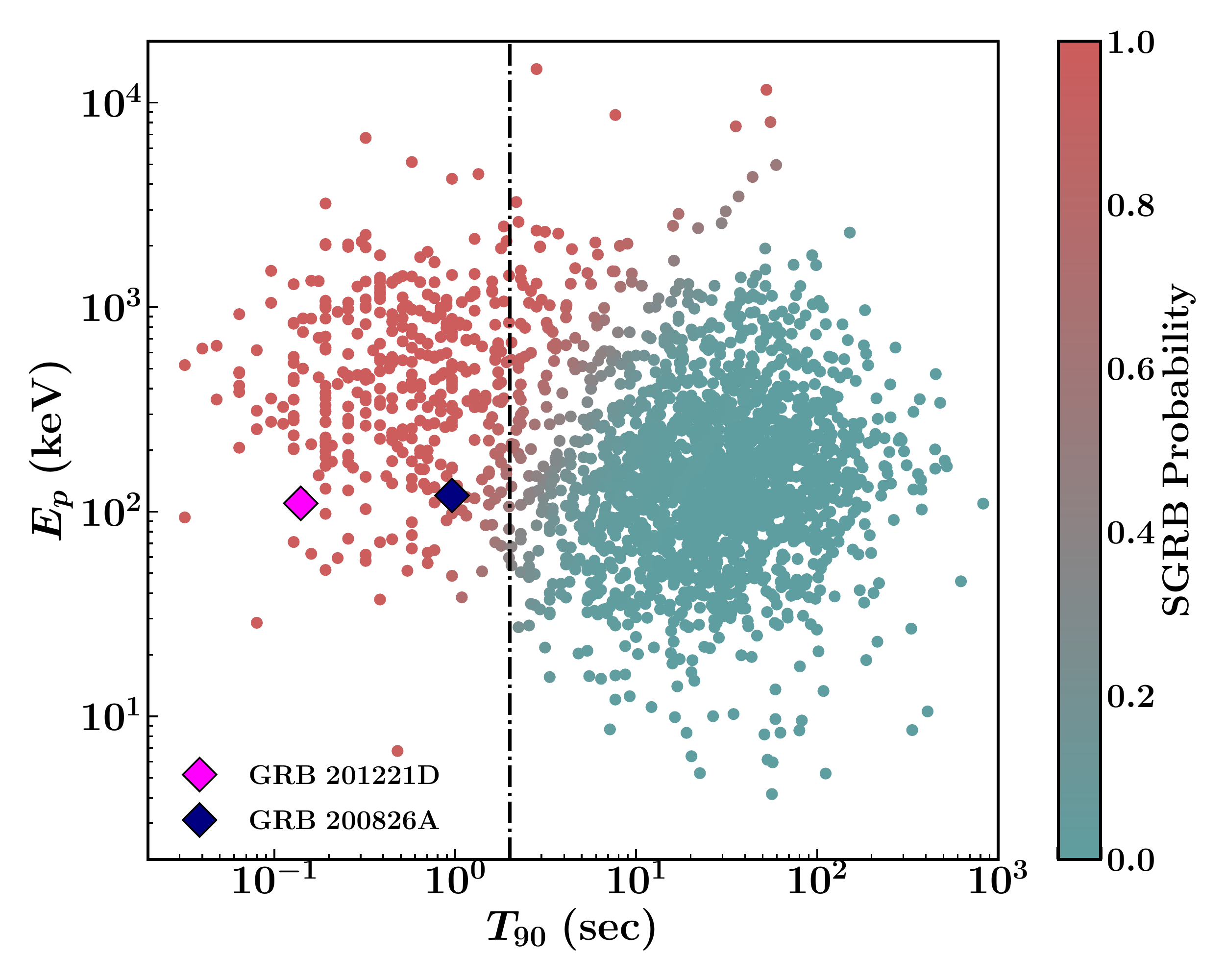}
\caption{The \Ep-\tninty distribution for GRBs taken from the \gbm catalogue. The magenta and blue diamonds indicate the location of \thisgrb and GRB 200826A in the distribution. The vertical line shows the traditional separation between short and long GRB at 2 sec. The colorbar on the right indicates the probability (estimated using BGMM) of GRBs being an SGRB.}
\label{ep-t90}
\end{figure}

\begin{figure}
\includegraphics[width=\columnwidth]{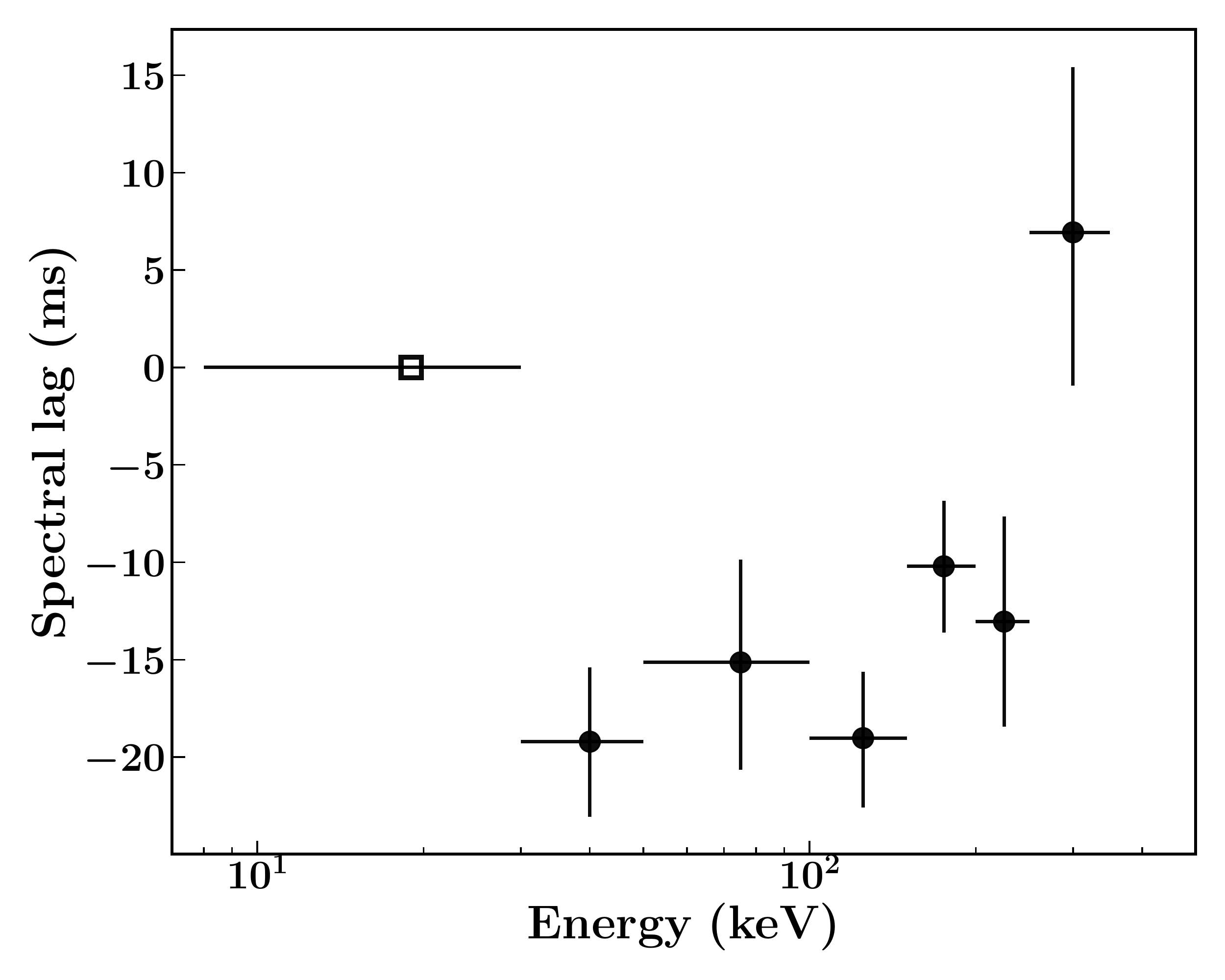}
\includegraphics[width=\columnwidth]{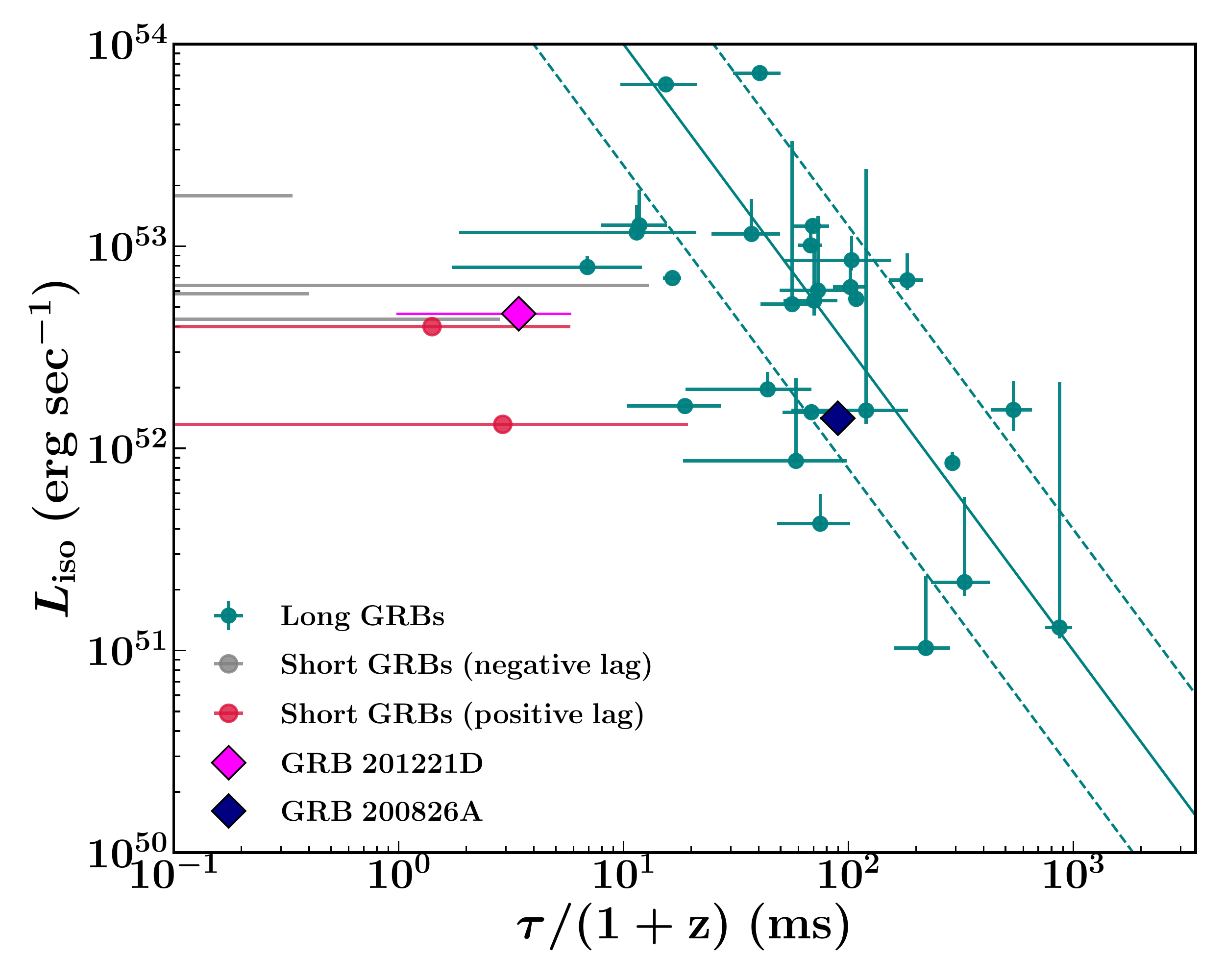}
\caption{{\bf Top:} The evolution of spectral lag for \thisgrb in different energy channels using the \fermi data. The value of lag is close to zero, as expected in the case of SGRBs. 
{\bf Bottom:} \thisgrb and GRB 200826A in the lag-luminosity plane. \thisgrb does not lie within the $2\sigma$ region (presented by dotted teal lines) of the lag-luminosity correlation. However, GRB 200826A follows this correlation, which is generally true for long GRBs.}
\label{spectral_lag}
\end{figure}

\subsubsection{Our observations}
\label{Dot_observations}
We observed the field of \thisgrb using the $4\rm K\times4\rm K$ ARIES Devasthal Faint Object Spectrograph and Camera  \citep[ADFOSC,][]{Omar2019} mounted on 3.6m DOT of ARIES Nainital. Four consecutive images of 15 min exposure time each were taken on December 22, 2020 ($\sim1$ day after the burst) in the $r^\prime$ band \citep{2020GCN.29148....1P}. The pre-processing of the images, including bias subtraction, flat-field correction and cosmic-ray removal, was performed using the \sw{Astropy} and \sw{CCDproc} modules in \sw{Python}. The cleaned images were aligned using \sw{astroalign} and stacked using the \sw{mediancombine} function of \sw{CCDProc} to improve the signal-to-noise ratio. An extended source is visible at the position of the burst (Fig. \ref{FC}). We performed \sw{PSF} photometry on the stacked image using \sw{DAOPHOT} and estimated the magnitude of the source to be $r^\prime=23.6\pm0.3$ mag (calibrated for the Pan-STARRS catalog). The late-time host galaxy observations were carried out on June 14, 2021, with the Hanle Faint Object Spectrograph and Camera (HFOSC) mounted on the 2.0m Himalayan Chandra Telescope (HCT). Four images of exposure time 900 sec each in the $R_C$ band were recorded. No source was detected to a magnitude limit of 22.9 mag (AB) in the stacked image. 

The field of \thisgrb was also observed with the TIFR-ARIES Near-Infrared Spectrograph (TANSPEC), one of the main instruments of 3.6m DOT. We took ten consecutive frames in the $R_C$ band with an exposure time of 500 sec each on June 4, 2021. The data pre-processing and photometry were performed in the same manner as described above. In the stacked image, we did not detect any source at the burst position to a magnitude limit of 23.2 mag (AB). 

\section{Results}
\label{results}

This section presents the results obtained from analysing the prompt emission of \thisgrb and its host galaxy properties. Due to the unavailability of sufficient X-ray and optical data, we 
could not perform an afterglow analysis.

\subsection {Spectral Hardness and Peak energy}
The hardness ratio (HR) is calculated using the ratio of counts in two energy channels (the $10-50$ keV and $50-300$ keV energy bands) for the selected three NaI detectors. The HR is estimated to be $2.68\pm0.83$, which is a typical value measured for SGRBs \citep[$3.61-5.64$ with a mean value of 4.61;][]{ohno_2008}. We plot the $E_{\rm p}$ - \tninty distribution for all GRBs taken from the GBM catalog \citep{von_2020}. As described in \S\ref{joint_spectrum} the value of $E_{\rm p}$ for \thisgrb was calculated by a joint \gbm and \bat spectral fit. We fit the $E_{\rm p}$ - \tninty distribution with a Bayesian Gaussian Mixture Model (BGMM), which is a machine-learning clustering algorithm generally used for classification. We find a probability of 98\% for \thisgrb to be a short burst. Fig. \ref{ep-t90} shows the $E_{\rm p}$ - \tninty distribution along with the probability of a GRB being short. The probability of GRB~200826A being an SGRB is 74\% \citep{zhang_2021}. However, recent analysis indicates a collapsar origin for GRB~200826A \citep{2021arXiv210505067A,2021arXiv210503829R} unlike SGRBs, which are proposed to come from compact object mergers. Even though the probability of \thisgrb belonging to the SGRB population is quite high, concerning the recent developments on GRB~200826A, we probe further to ascertain the classification of GRB~201221D.

\begin{table}
\centering
\caption{Spectral lag of \thisgrb in different energy channels with reference to the $8-30$ keV band.}
\begin{tabular}{ccl}
\hline
\bf {Energy Channel \ (\keV)} & & \bf {Spectal lag (ms)} \\
\hline 
30--50  & & $-19.2_{-3.85}^{+3.82}$ \\ \T
50--100  &  &$-15.1_{-5.50}^{+5.27}$\\ \T
100--150  & &$-19.0_{-3.55}^{+3.41}$ \\ \T
150--200  & &$-10.2_{-3.40}^{+3.37}$ \\ \T
200-250  & &$-13.0_{-5.39}^{+5.40}$ \\ \T
250--350  & & $+6.93_{-7.86}^{+8.49}$ \T \\

\hline

\end{tabular}
\label{tab:spectral lag}
\end{table}

\subsection{Spectral lag}
We calculate the spectral lag for \thisgrb in different energy bands, selecting the range between $8-350$ \keV (a sufficient number of counts are not available beyond 350 keV), considering the $8-30$ \keV band as the reference channel. We estimate the temporal correlation of the two light curves using the cross-correlation function (CCF) as described in \citet{Bernardini2015}. The maximum of the temporal correlation provides the delay between two light curves. To find the global maximum, we fit the correlation with an asymmetric Gaussian function using \sw{emcee} \citep{emcee2013}. The spectral lag in different energy bands are quoted in Table \ref{tab:spectral lag} and the evolution is shown in the top panel of Fig. \ref{spectral_lag}. 

An anti-correlation has been found between the bolometric peak luminosity and the spectral lag of GRBs by \citet{norris2000}, later confirmed by \citet{norris2002,Gehrels2006,Ukwatta_2010}. To put \thisgrb in lag-luminosity correlation, we calculate the lag between the two energy channels ($15-25$ \keV and $50-100$ keV) of \bat to compare (the same energy channels used for the sample of GRBs defined in \citet{Ukwatta_2010}). The lag between the BAT energy channels is $7\pm5$ ms, close to zero within errors. 

The burst does not lie within the $2\sigma$ region of the lag-luminosity correlation, as shown in the bottom panel of Fig. \ref{spectral_lag}. On the other hand, the lag measured in GRB~200826A was 157 ms \citep{zhang_2021}, and it falls within the lag-luminosity correlation, which is generally true for long GRBs. It increases the ambiguity in the classification of GRB~200826A.

\subsection{Non--Collapsar Probability}
As discussed earlier, the origin of SGRBs belongs to old stellar populations and is supposed to lie at low redshifts \citep{ 2010ApJ...725.1202L, Fong2013}, but \thisgrb lies at a high redshift ($z=1.045$) as compared to the median redshift of SGRBs. Therefore, to check if \thisgrb originated from a collapsar or not, we estimate the non-collapsar probability ($f_{nc}$) using the functions defined in \citet{Bromberg_2012, Bromberg2013}:
\begin{equation}
\label{eq:fNC}
f(T_{90})=A_{NC}\frac{1}{T_{90}\sigma\sqrt{2\pi}}e^{-\frac{(\ln T_{90}-\mu)^2}{2\sigma^2}}\left(\frac{dN_{GRB}}{dT_{90}}\right)^{-1},
\end{equation}

where, $dN_{GRB}/dT_{90}$ represents the non-Collapsar distribution and is given by equation: 

\begin{equation}
\label{eq.dNdT.GRB}
\begin{aligned}
\frac{dN_{GRB}}{dT_{90}}=A_{NC}\frac{1}{T_{90}\sigma\sqrt{2\pi}}e^{-\frac{(\ln T_{90}-\mu)^2}{2\sigma^2}} 
\\ +
A_C  \left\{
\begin{array}{lc}
1 & T_{90}\leq T_B \\
\left(\frac{T_{90}}{T_B}\right)^\alpha e^{-\beta(T_{90}-T_B)} & T_{90}>T_B,
\end{array}
\right.
\end{aligned}
\end{equation}

The first and the second term correspond to non-collapsars and collapsars, respectively. 
$T_{B}$ is the observed breakout time in the duration distribution. $A_{NC}$ and $A_{C}$ are the fit parameters and are taken from \citet{Bromberg2013} that they obtained by fitting the duration distributions to the collapsar distribution function.

Using \tninty (\gbm) for GRB~201221D, we estimate the $f_{nc}$ value of $0.95\pm0.09$. For comparison, we also calculate the $f_{nc}$ value for GRB~200826A, which is $0.70\pm0.01$. The high probability of a non-collapsar origin for \thisgrb shows that it very likely belongs to the non-collapsar progenitors.

\subsection{Host Properties}

\begin{figure*}
\centering
\includegraphics[width=\linewidth]{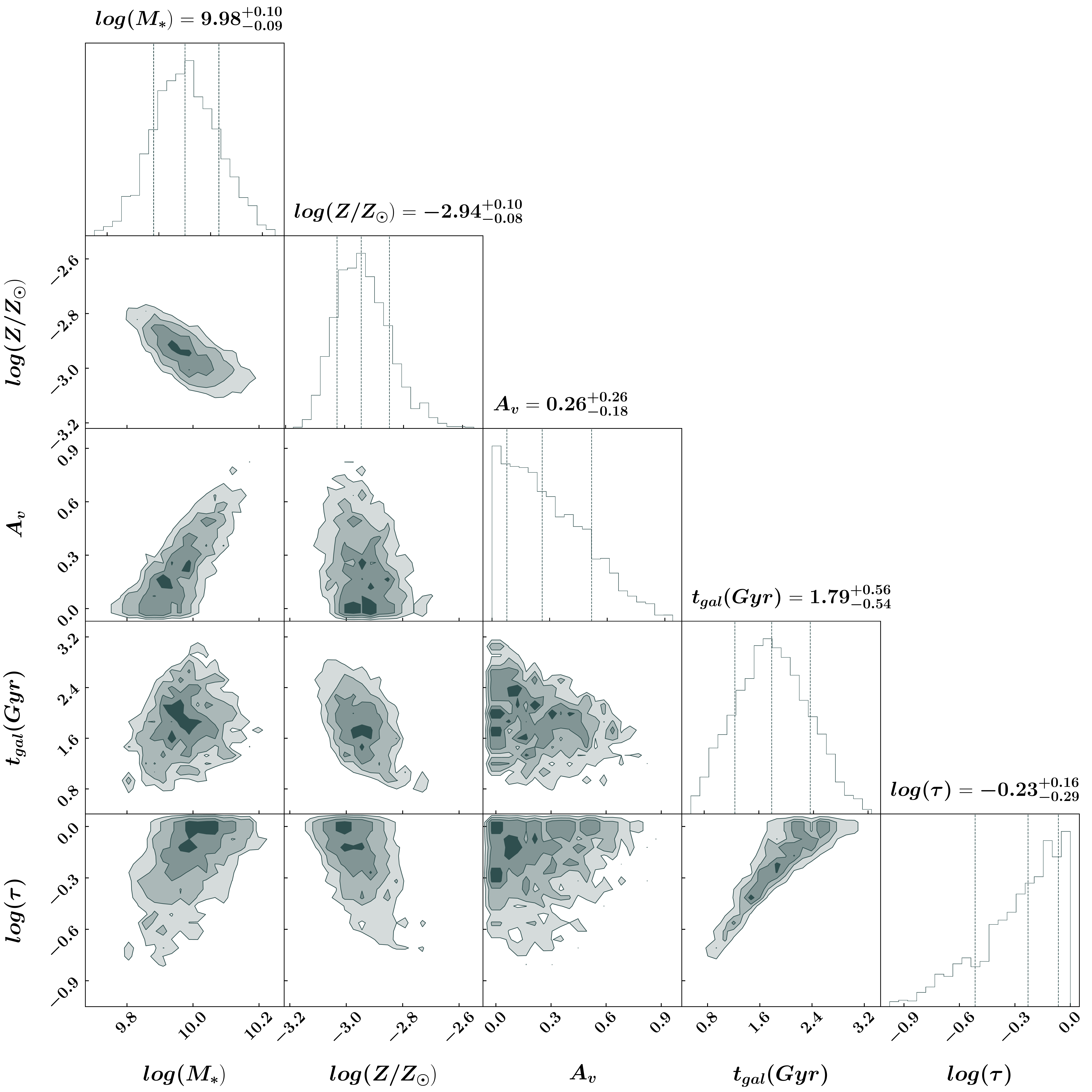}
\caption{ The posterior distributions for various host parameters obtained from \sw{Prospector}.}
\label{prospector_corner}
\end{figure*}

\begin{figure}
\centering
\includegraphics[width=\columnwidth]{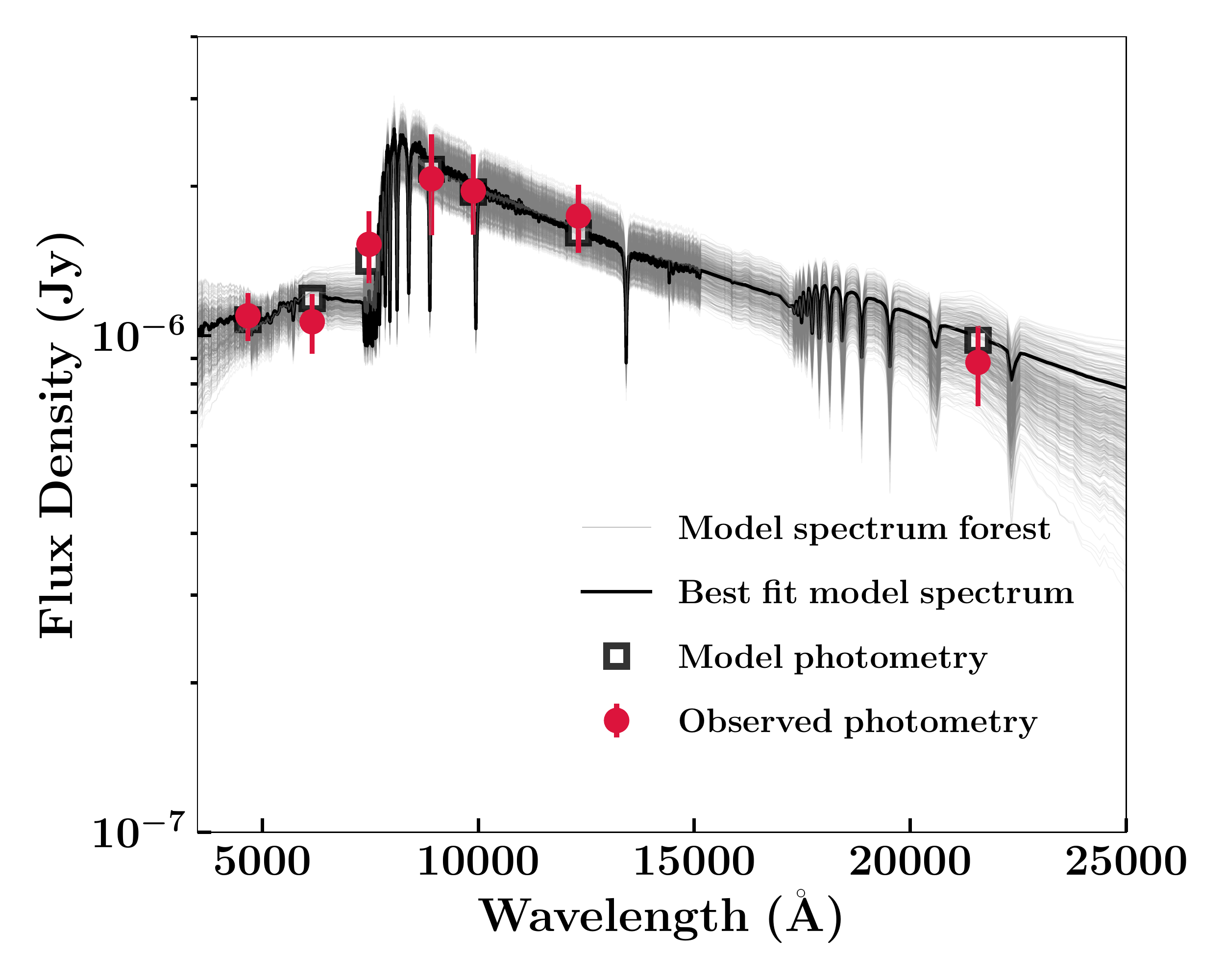}
\caption{ The host SED along with the 500 forest plots (light grey) which are randomly generated from the posterior distributions of the parameters obtained from \sw{Prospector}. The black open squares and the thick black plot represents the best fit model photometry and best fit spectrum, respectively. The model photometry and spectrum are in agreement with the observed data.}
\label{prospector_sed}
\end{figure}

The host galaxy of \thisgrb was identified in the optical and NIR bands with LBT in late-time observations \citep{fernandez2021grb}. The host magnitudes are listed in Table \ref{tab:optical_data}. The available magnitudes are used to investigate the host galaxy properties using \sw{Prospector} \citep{Johnson2021ApJS}. \sw{Prospector} is a \sw{Python}-based stellar population modelling code which uses Flexible Stellar Population synthesis \citep[FSPS;][]{Conroy2009, Conroy_2010, Conroy2013} to build the stellar population models \citep{Leja2017, Johnson2021ApJS}. It utilises \sw{Dynesty} \citep{Speagle_2020}, a nested sampling algorithm, to fit the photometric and spectroscopic data of a galaxy and provides the best-fit solution and posterior parameter distributions for the galaxy parameters. We used the best fit to determine the stellar mass ($M$), age of the galaxy ($t_{gal}$), star-formation history (SFH), dust extinction ($A_V$), and stellar metallicity ($Z$) using the methodology described in \citet{Johnson2021ApJS}. We used the milky way extinction law and Chabrier initial mass function \citep{Cardelli_1989,Chabrier_2003} . We fixed the redshift to $z=1.045$ \citep{fernandez2021grb} and fitted for other parameters by setting the priors as listed in Table \ref{tab:host_properties}. The maximum value of the age of the galaxy is fixed to 6.148~Gyr, the age of the Universe at the redshift of the burst. The posterior distributions for the parameters produced using \sw{Prospector} are shown in the corner plots in Fig. \ref{prospector_corner}. The photometric data of the host over-plotted with the model spectrum and photometry is shown in Fig. \ref{prospector_sed}. We found the best-fit values for the parameters as listed in Table \ref{tab:host_properties} with log evidence value of $148\pm36$. The values of the host parameters are consistent with the values derived in \citet{fernandez2021grb}.

Further, we estimated the SFR using the relation:
\begin{equation*}
\text{SFR}(t) = M\times \Big[\int_0^t{te^{-t/\tau} dt}\Big]^{-1} \times te^{-t/\tau}, 
\end{equation*} 
where, $M$ is the total mass of the galaxy, $t$ is the age of the galaxy and $\tau$ is star-formation timescale. 
The value of SFR is given in Table \ref{tab:host_properties}. This relatively high value of SFR is consistent with the detection of O[II] emission from the host in the GTC spectrum.

\begin{table}
\centering
\caption{Host properties of \thisgrb estimated from SED fitting using \sw{Prospector}.}
\begin{tabular}{l c r}
\hline
\bf {Host Properties} & \bf{Priors} & \bf {Values} \\
\hline \T
log$(M_{*}$) & 9.0 -- 11.0 & $9.98_{-0.09}^{+0.10}$\\\T   
log$(Z/Z_{\odot}$) & -4.0 -- 4.0 & $-2.94_{-0.08}^{+0.10}$\\\T  
$A_V (mag)$ & 0 -- 2.0 &$0.26_{-0.18}^{+0.26}$\\\T 
t$_{gal}(Gyr)$ & 0 -- 6.1  & $1.79_{-0.54}^{+0.56}$\\ \T
SFR($M_{\odot}yr^{-1}$) & -- &$2.92\pm1.43$\\ 
\hline
\end{tabular}
\label{tab:host_properties}
\end{table}

\begin{figure}
\includegraphics[width=\columnwidth]{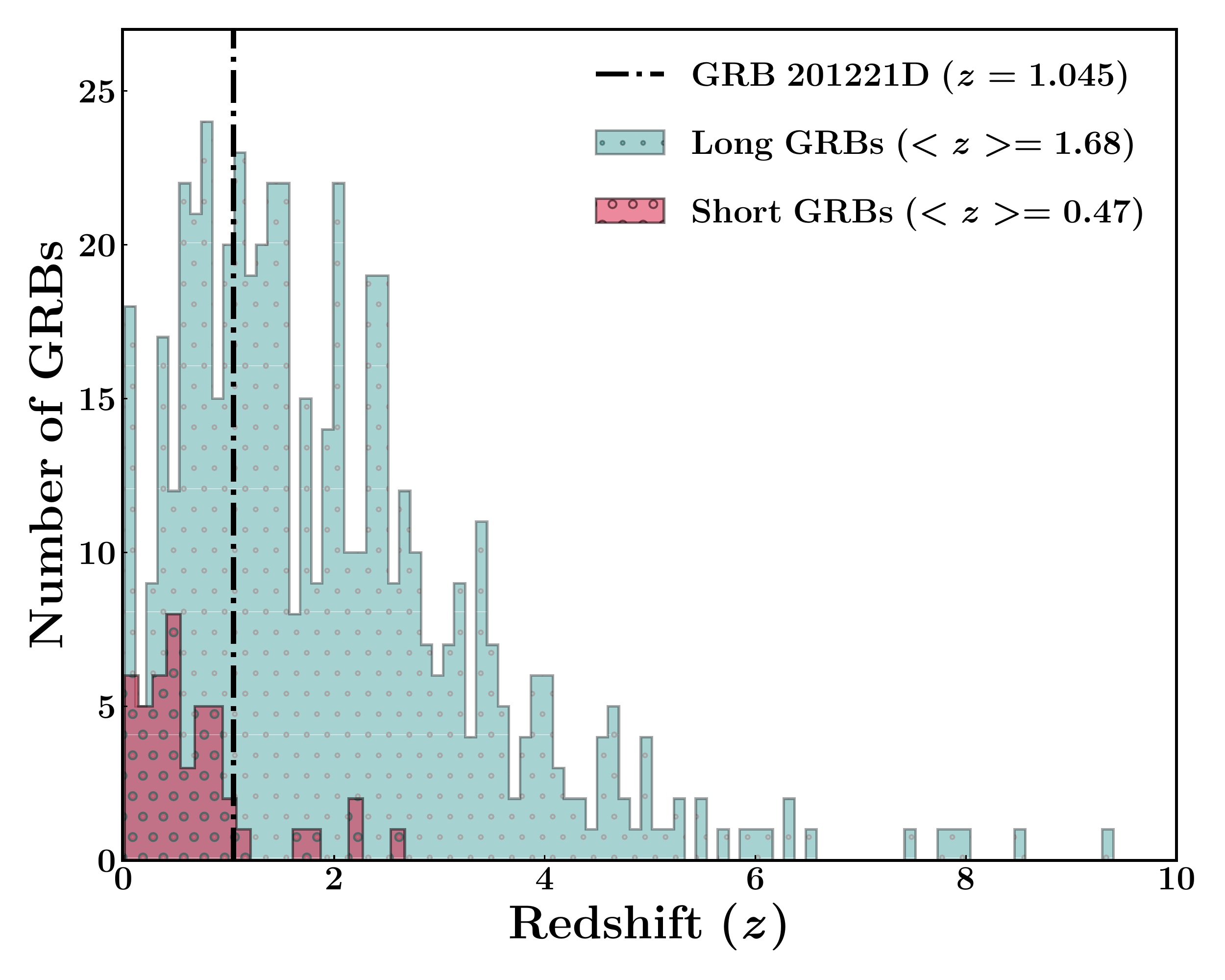}
\caption{Redshift distribution of GRBs (both long and short) up to October 2021 (from \citealt{M2019} and Jochen Greiner's compilation page). The SGRBs lie at the lower end of the redshift distribution with a median redshift of 0.47. Long GRBs are spread across the redshift distribution with a median of 1.68.}
\label{redshift}
\end{figure}

\begin{figure}
\includegraphics[width=\columnwidth]{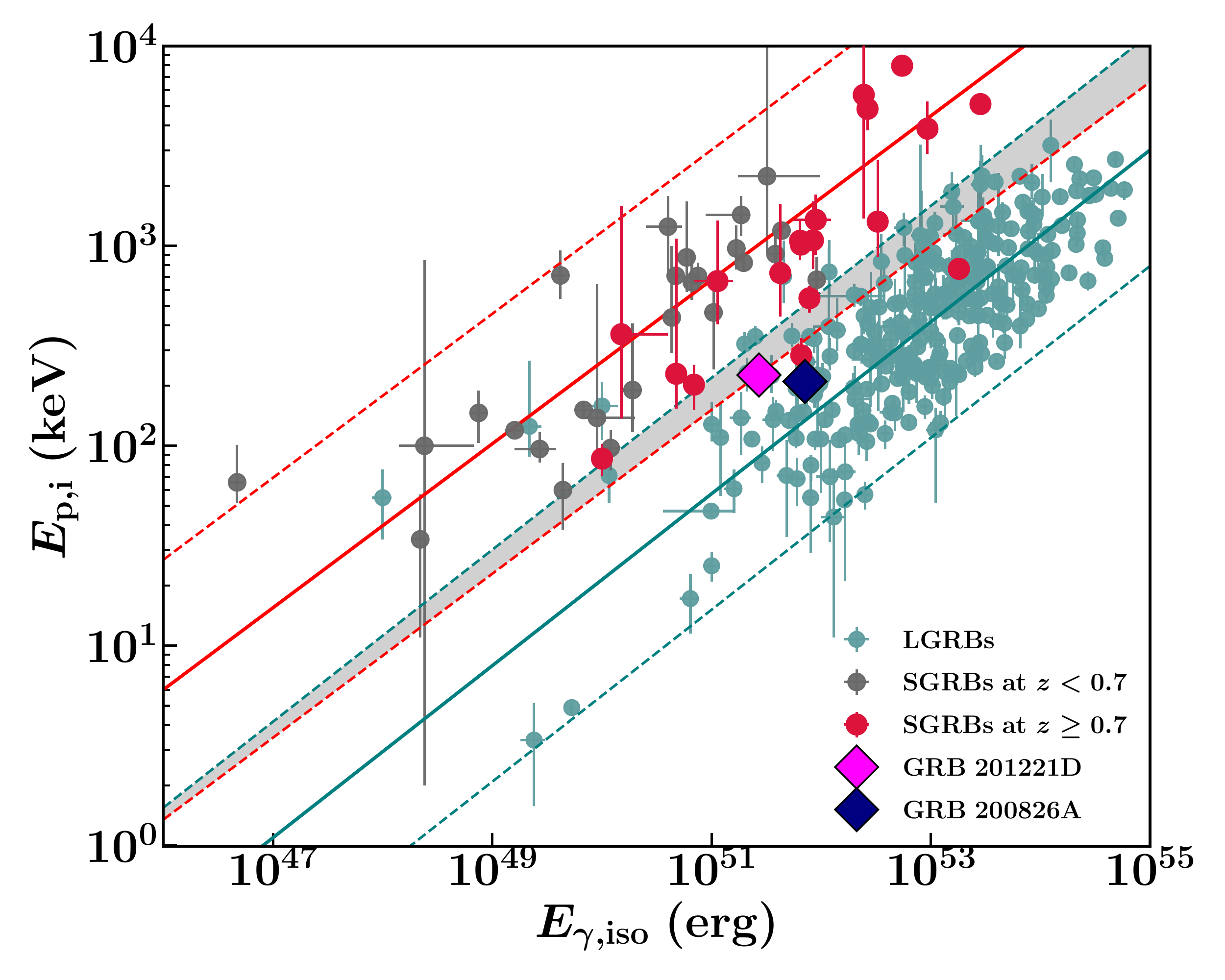}
\includegraphics[width=\columnwidth]{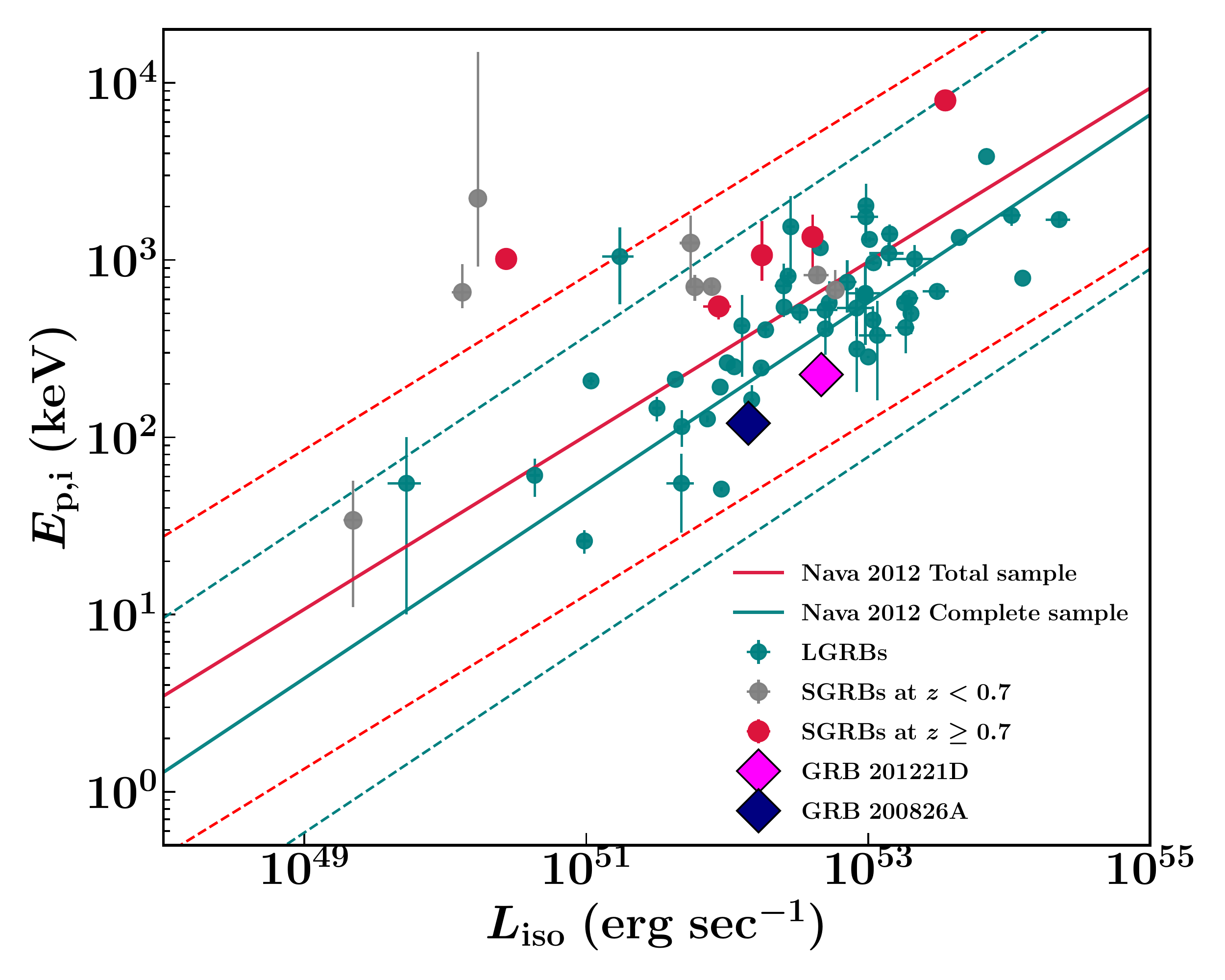}
\caption{ {\bf Top:} Short and long GRBs in the Amati correlation plane. The SGRBs are divided into Group 1 (at $z<0.7$, grey circles) and Group 2 ($z\geq0.7$, crimson circles). The locations of \thisgrb and GRB~200826A in the plane are shown with magenta and blue diamond symbols, respectively. GRB~200826A, along with some other SGRBs from Group 2 are seen to follow the track of long GRBs. However, \thisgrb lies in the overlapping region of $2\sigma$ tracks of long and SGRBs. {\bf Bottom:} Yonetoku correlation for Group 1, Group 2 SGRBs along with long GRBs. Both \thisgrb and GRB~200826A follow the Yonetoku correlation and are situated well within the $2\sigma$ region of the correlation.}
\label{correlations}
\end{figure}

\section{Are high redshift SGRBs similar to low redshift SGRBs?}
\label{comparison}

Motivated by the fact that some of the SGRBs lying at a high redshift (e.g., GRBs 200826A and 090426A) have signatures of collapsars and earlier prediction by \citet{Berger2007} that there can be a new population of SGRBs at higher redshifts, we examine the similarity and differences between SGRBs at low and high redshifts. 

We selected all the GRBs (both long and short available in Jochen Greiner's compilation page\footnote{\url{https://www.mpe.mpg.de/~jcg/grbgen.html}}) up to October 2021 with known redshifts and calculated the redshift corrected \tninty {$T_{\rm 90,i}$}. We selected all the GRBs with $T_{\rm 90,i} < 2$ sec as the SGRBs in our sample. The full sample of 43 SGRBs is given in Table \ref{tab:sample}. 
We compared the redshift distribution of SGRBs with that of long GRBs. Fig. \ref{redshift} shows the redshift distribution of GRBs. We estimated the median redshift value for SGRBs is $\overline{z}=0.47$, which is lower than the estimated median redshift value of long GRBs ($\overline{z}=1.68$). Considering the median redshift of SGRBs, We divide the sample of SGRBs into two groups; Group 1-low redshift SGRBs with $z<0.7$, and Group 2-high redshift SGRBs with $z\geq0.7$. 

In this section, we present the comparison of prompt (prompt emission correlations and $f_{nc}$) properties, afterglow, and the host properties of SGRBs lying at high and low redshifts.

\begin{figure}
\includegraphics[width=\columnwidth]{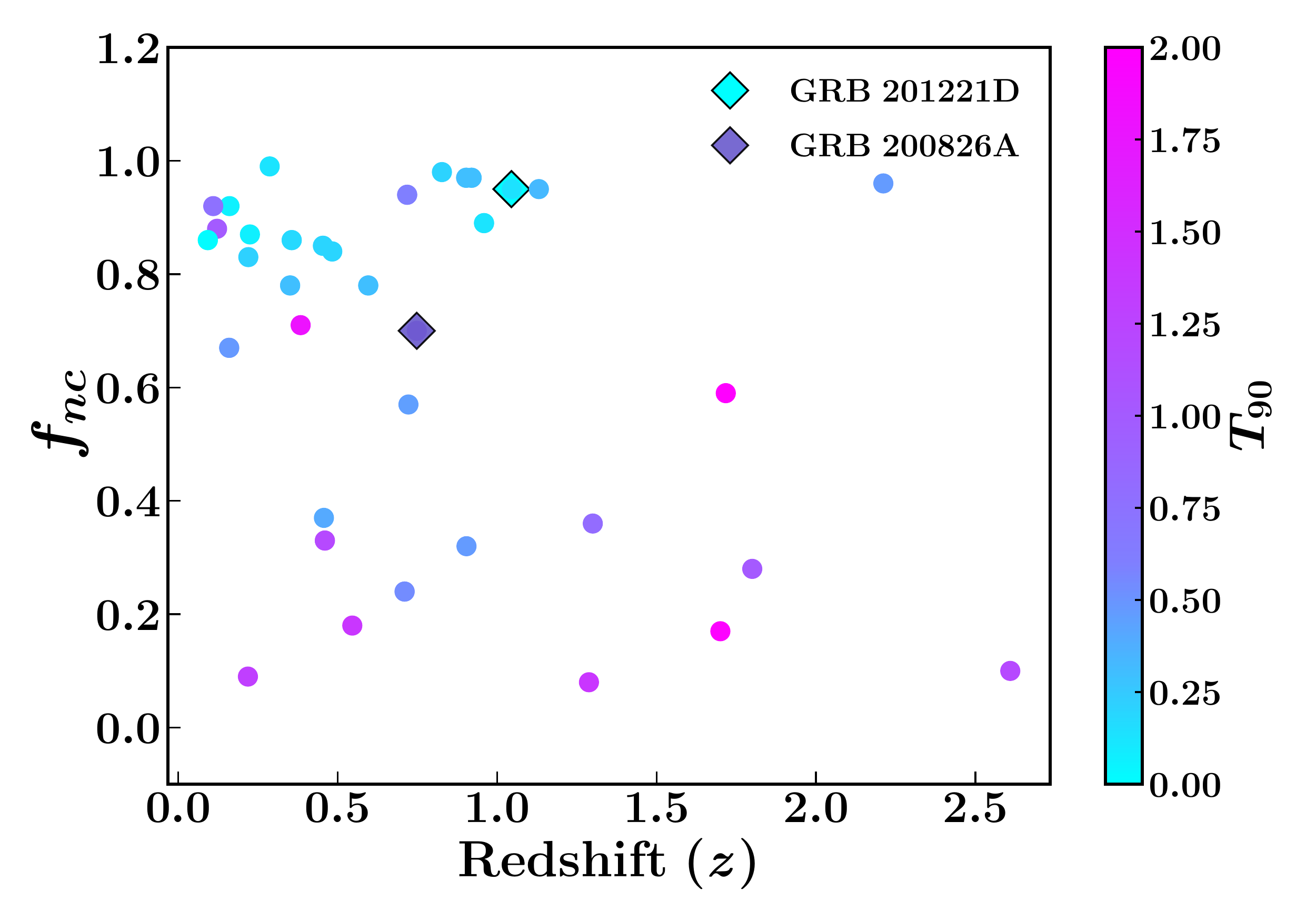}
\caption{Non-collapsar probability of SGRBs along with their redshifts. The colorbar indicates the \tninty values of SGRBs. The $f_{nc}$ value is 0.4 for most of the SGRBs lying at $z>1$, indicating that some of these SGRBs might have originated from collapsars.}
\label{fnc_comp}
\end{figure}

\subsection{ {Prompt emission properties}}
Prompt emission correlations have been used as tools to classify GRBs for a long time. In the Amati correlation plane, $E_{\rm \gamma, iso}$ - $E_{\rm p,i}$ (peak energy in the source frame) plane, two classes of GRBs lie at different positions following different tracks \citep{Amati_2002, Amati_2006}. Using the fluence value (1-1000 keV) and $E_{\rm p}$ values estimated in section \ref{joint_spectrum}, we calculate the isotropic energy release and $E_{\rm p,i}$ for GRB~201221D, $E_{\rm \gamma, iso}=2.762\times10^{51}$ erg and $E_{\rm p,i} = 226_{-35.8}^{+31.8}$. 


We plot \thisgrb in the Amati correlation plane along with SGRBs of Group 1 and Group 2 and long GRBs (Fig. \ref{correlations}) using the values of $E_{\rm \gamma, iso}$, $E_{\rm p,i}$ from \cite{M2019}. The dotted lines show the $2\sigma$ correlation regions. \thisgrb lies in the overlapping $2\sigma$ regions of correlation of both short and long GRBs. In the figure, we also highlight the position of GRB~200826A, which follows the long GRB track. We also find that some SGRBs at $z>0.7$ lie on the long GRB track and some in the overlapping $2\sigma$ correlation region of short and long GRBs. 

In addition, we calculate $L_{\rm \gamma,p, iso} = 4.64\pm0.84\times10^{52}$ and put \thisgrb in the Yonetoku correlation plane ($L_{\rm \gamma,p, iso}$-$E_{\rm p, i}$) along with the sample available from the literature \citep{Yonetoku_2004,2012MNRAS.421.1256N}. The bottom panel of Fig. \ref{correlations} shows the location of the burst in the Yonetoku plane, \thisgrb and GRB~200826A lie within the $3\sigma$ scatter of the sample of GRBs studied by \citet{2012MNRAS.421.1256N}.

We also compare the non-collapsar probability ($f_{nc}$) of \thisgrb and GRB~200826A with other SGRBs with a known redshift from the sample of \cite{Bromberg2013}. Fig. \ref{fnc_comp} shows the $f_{nc}$ for SGRBs lying at different redshifts. These results indicate that most of the SGRBs at high redshift ($z>1$) have lower values of $f_{nc}$, which is in agreement with the results of \citet{Bromberg2013}, indicating that these SGRBs might arise from progenitors other than compact object mergers.

\begin{figure}
\centering
\includegraphics[width=\columnwidth]{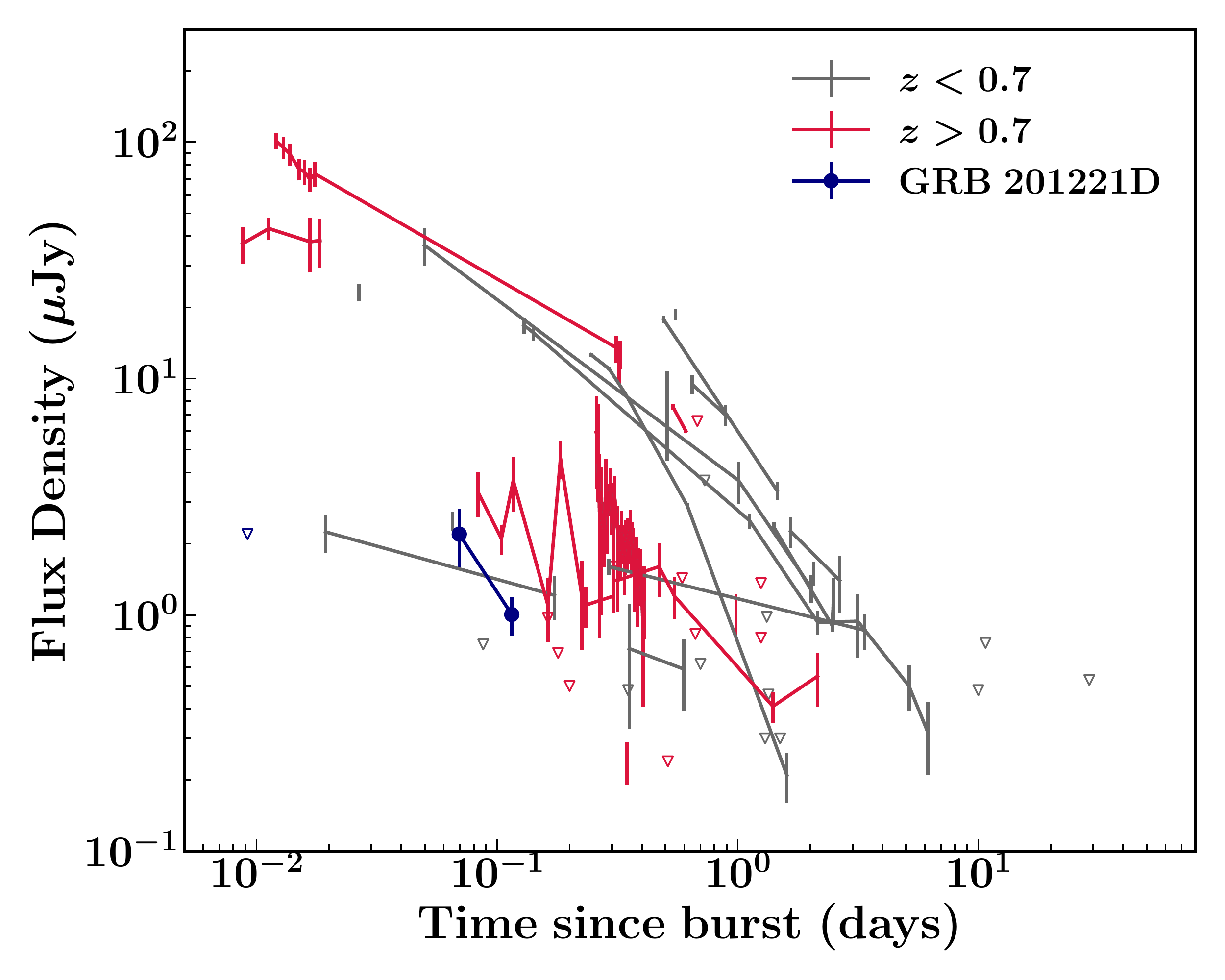}
\includegraphics[width=\columnwidth]{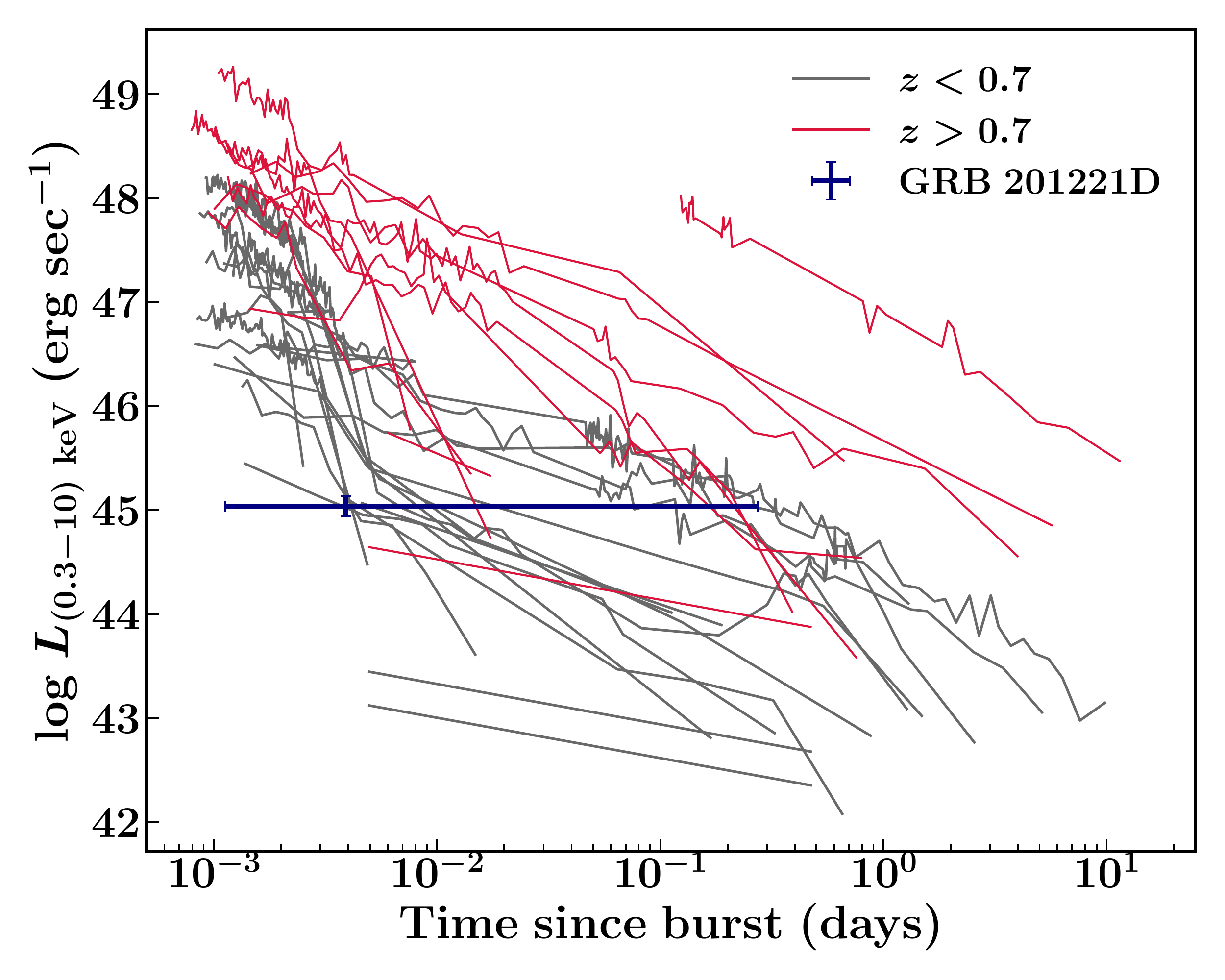}
\caption{Optical and X-ray light curves of SGRBs. The sample of SGRBs is divided into Group 1 ($z< 0.7$) and Group 2 ($z\geq0.7$) as indicated in the plot. The light curves of \thisgrb are shown in blue color.}
\label{op_xray_lcs}
\end{figure}

\subsection{Multi-band SGRB afterglow light curves}
We compare the optical ($R_C/r^\prime$) and X-ray (0.3 - 10 keV) afterglow light curves of Group 1 and Group 2 SGRBs, as defined earlier. We construct the optical light curves of SGRB afterglows using the data from \cite{Fong2015} up to 2015 and \cite{Rastinejad2021} for bursts beyond 2015. The magnitudes are converted to flux density after correcting for galactic extinction for each burst.   

The X-ray light curves, in units of flux, in the energy range $0.3-10$ keV, are taken from the \swift XRT repository\footnote{\url{https://www.swift.ac.uk/xrt_curves/}}. The flux light curves are converted to luminosity to compare Group 1 and Group 2 SGRBs. Fig. \ref{op_xray_lcs} shows the comparison between optical and X-ray light curves of Group 1 and Group 2 SGRBs. The optical light curves show a wide range in brightness for SGRBs at different redshifts. However, the X-ray luminosities for SGRBs at high redshifts (Group 1) are systematically higher than those of SGRBs at lower redshifts. However, It can also be because the faint bursts would not be detected at high-redshifts, where only luminous bursts can be detectable.

\thisgrb does not have good coverage in both optical and X-ray bands. With the limited data \thisgrb seems to lie at the lower end of the luminosity distribution in optical and X-ray bands.

\subsection{Host Properties}

\begin{figure}
\includegraphics[width=\columnwidth]{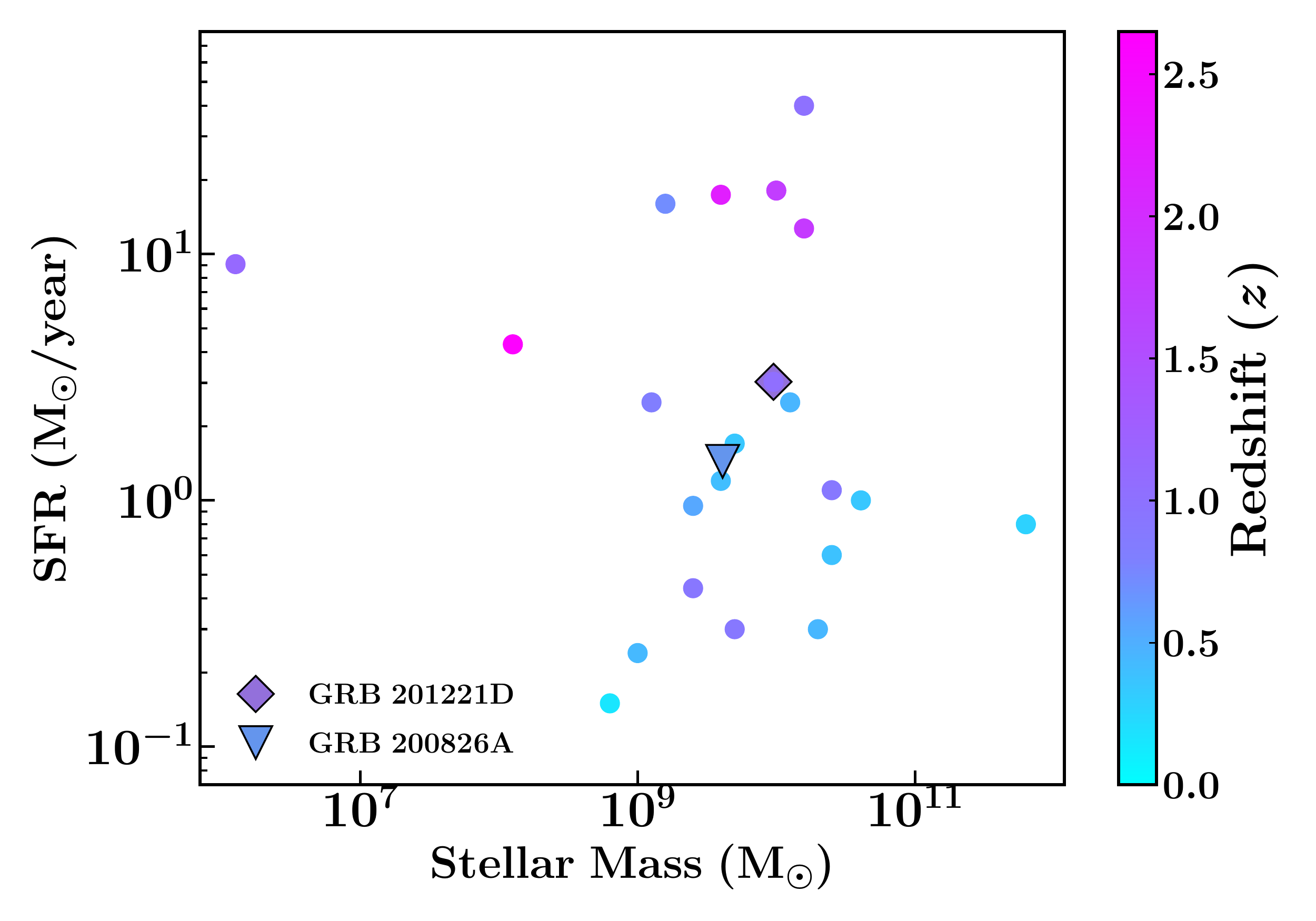}
\caption{Star-formation rate and stellar mass of SGRB hosts color-coded with the redshift value. The data are taken from \citet{Berger2014, Dichiara_2021, zhang_2021} and this work.
The SGRB hosts lying at higher redshifts have relatively larger SFR values than those for low-redshift SGRB hosts. \thisgrb has an intermediate value compared to the sample of SGRBs studied here.}

\label{host_comp}
\end{figure}

We compare the SFRs of the hosts of all SGRBs with known redshifts. The SFR values are taken from \cite{Berger2014} and \cite{Dichiara_2021}. In Fig. \ref{host_comp} we plot the SFR and stellar mass of all SGRB hosts along with \thisgrb (this work) and GRB~200826A \citep{zhang_2021} color-coded with the redshift value. We notice that the hosts of SGRBs lying at higher redshifts have higher SFR values than those of SGRBs lying at lower redshifts. A recent study by \cite{Dichiara_2021} compared the SFRs of SGRB hosts at redshift $z>1$ with those of long GRBs at redshift $1<z<2$. Their study indicated a significant overlap in SFR and stellar masses between short and long GRB hosts in this redshift range.

As SGRBs are supposed to originate from compact star mergers, they are believed to be associated with an old population of galaxies with low SFRs \citep{Fong_2013,Berger2014,Li_2016}. On the other hand, long GRBs, expected to originate from massive star collapsars, are generally found in star-forming galaxies with high SFRs. 
The overlap between the SFRs of long and SGRB hosts at redshift $z>1$ indicates they might have the same type of progenitor system. However, we can not deny the fact that galaxy properties also vary with redshift. In general, there is a steady decrease in the overall SFR of the Universe by a factor of
10 from to z=1 to z=0 \citep{Madau_1996,Bauer_2005}.

\section{Summary}
\label{conclusions}

We have presented the analysis of \thisgrb and its comparison with the SGRB sample. We determined the prompt emission parameters such as spectral hardness, lag, non-collapsar probability and the host galaxy properties of the burst. We also performed the time-resolved spectroscopy of the prompt emission of \thisgrb and compared the evolution with that of an SGRB sample from \cite{burgess2017bayesian}. The fit parameters, $E_{\rm p}$ (peak energy), $\alpha$ (spectral index) and flux show hard-to-soft evolution. The ($\alpha$ and flux) lie well within the usual range for the SGRB sample. The $E_{\rm p}$ value is softer than the SGRB sample and comparable to that of long GRBs. 

As the \tninty value depends on the sensitivity of an instrument and the background variations, it alone can not decide the classification of a burst . It is essential to look for other properties that can be used for classification \citep{2000PASJ...52..759Q, 1995ApJ...448L.101F}.
We used different methods reported in the literature to confirm the class of GRB~201221D \citep{M2019, Dimple_2022}. 
We calculated the probability of \thisgrb being an SGRB by fitting the \Ep-\tninty distribution with BGMM. The probability of \thisgrb is 98\%, indicating that \thisgrb very likely belongs to SGRB class. Furthermore, we calculated the spectral lag in different energy bands, and the lag value is close to zero, as expected for SGRBs. We placed the burst in the Amati correlation plane. It lies in the overlapping region of short and long GRBs.

Furthermore, we compared the prompt (prompt emission correlation and $f_{nc}$), afterglow, and host properties of SGRBs lying at high and low redshifts to address the implication of redshift on the progenitor system of SGRBs. We found that:

(a) SGRBs with $z>0.7$ are located close to the long GRB track in the Amati plane. Three SGRBs (including GRB 200826A) lie on the long GRB track. Some of these SGRBs, including \thisgrb lie in the overlapping region of $2\sigma$ regions of long and SGRBs. 

(b) The non-collapsar probabilities for some high redshift SGRBs have values $<0.5$, indicating these SGRBs might result from collapsars. 

(c) The optical brightness covers a wide range for SGRBs at different redshifts. However, the X-ray luminosities for SGRBs at high redshifts are systematically higher than that of SGRBs at lower redshifts. Also, a fraction of high redshift SGRB hosts has large SFRs comparable to those of long GRB hosts. However, this can be an observational artefact.

The studies show that SGRBs lying at high redshifts have similarities to long GRBs, indicating they might have progenitor systems other than compact object mergers (e.g. GRB~200826A and GRB~090426), or there might exist subgroups within the SGRBs originating through different channels \citep{Yu_2018}. The investigation of SGRBs lying in the overlapping region can provide a clearer picture of the progenitor systems of SGRBs.

Machine learning algorithms can play a crucial role to solve the classification conundrum \citep{Christian_2020,Dimple_2022}. In addition, late-time optical and NIR observations in the future can help to observe the bumps in the optical/NIR light curves in GRBs. It will lead to identifying the bumps as supernovae/kilonovae, which are uniquely associated with collapsars/compact binary mergers. However, it is difficult to detect the kilonova/supernova transients at higher redshifts due to the observational limitations and their faintness and fast evolution. The observations by future telescopes like Extremely Large Telescope (ELT), Thirty-Meter Telescope (TMT), and Giant Magellan Telescope (GMT) have the potential to detect kilonovae at redshift > 1. In addition to optical observations, gravitational-wave observations have immense potential to shed light on this problem. However, only third-generation gravitational wave detectors such as Einstein Telescope and Cosmic Explorer~\citep{2012CQGra..29l4013S, evans2021horizon,kalogera2021generation} expected to be operational in 2030+, will have the sensitivity to observe binary neutron stars at redshifts around 1. 

\section*{Acknowledgements}

We thank the referee for the constructive comments which has improved the presentation of the article. KM, RG, and SBP acknowledge the BRICS grant {DST/IMRCD/BRICS/PilotCall1/ProFCheap/2017(G)} for the financial support. KGA is partially supported by the Swarnajayanti Fellowship Grant No.DST/SJF/PSA-01/2017-18, MATRICS grant MTR/2020/000177 of SERB, and a grant from the Infosys Foundation. This research is based on observations obtained from the 3.6m DOT during observing cycle DOT-2020-C2, a National Facility run and managed by ARIES, an autonomous institute under the Department of Science and Technology (DST), Government of India. This research has used data obtained from the High Energy Astrophysics Science Archive Research Center (HEASARC) and the Leicester Database and Archive Service (LEDAS), provided by NASA's Goddard Space Flight Center and the Department of Physics and Astronomy, Leicester University, UK, respectively.

\section*{Data Availibilty}
The optical data is already presented in the article, and other data sets are available in the public domain. 


\bibliographystyle{mnras}
\bibliography{main} 


\defcitealias{Bromberg2013}{1}
\defcitealias{Berger2014}{2}
\defcitealias{Dichiara_2021}{3}
\defcitealias{host_140903}{4}
\defcitealias{host_150101b}{5}
\defcitealias{host_170817A}{6}
\defcitealias{zhang_2021}{7}
\onecolumn
\begin{longtable}{cccccccc}
\caption{\label{tab:sample} Sample of SGRBs with known redshifts}\\ \hline \T
%
& & &  &Group 1 SGRBs  & & \T \\
\\
\hline
\\
GRB  & $T_{\rm 90,i}^{a,b}$ & ${z}^{a}$ & $E_{\rm \gamma, iso}^{a}$ & $E_{\rm p,i}^{a}$ & $f_{nc}$ & SFR & References$^{c}$ \\ \T
 & (sec) & & $10^{51}$ \ (erg) & (\keV) & & ($\rm M_{\odot}/year$) \\
\\
\hline
\endfirsthead
\caption{continued...}\\
\hline
\T
GRB  & $T_{\rm 90,i}^{a,b}$ & ${z}^{a}$ & $E_{\rm \gamma, iso}^{a}$ & $E_{\rm p,i}^{a}$ & $f_{nc}$ & SFR & References$^{c}$ \\ \T
 & (sec) & & $10^{51}$ \ (erg) & (\keV) & & ($\rm M_{\odot}/year$) \\
\\
\hline  
\endhead
\hline
\endfoot
\vspace{1mm}

GRB~050509B  &  0.04 & 0.2248 &  $0.0024_{-0.001}^{+0.004}$  & $100_{-98}^{+748}$ & $0.87_{-0.16}^{+0.04}$ & <0.15 & \citetalias{Bromberg2013,Berger2014} \T \\    
GRB~050709  &  0.06  &  0.1606  &  $0.027_{-0.011}^{+0.011}$ & $96.3_{-13.9}^{+20.9}$ &  ${0.92^{+0.02}_{-0.03}}$ & 0.15 & \citetalias{Bromberg2013,Berger2014}\T \\ 
GRB~050724 &  2.4 & 0.2576 & $0.090_{-0.02}^{+0.11}$ & $138_{-57}^{+503}$ & -- & <0.1 & \citetalias{Berger2014}\T \\
GRB~051221A  &  0.14 & 0.5465   &  $9.10_{-1.12}^{+1.29}$ &  $677_{-141}^{+200}$& $0.18^{+0.08}_{-0.11}$ &0.95 & \citetalias{Bromberg2013,Berger2014} \T \\    
GRB~060502B  &  0.12  &  0.287   & $0.433_{-0.053}^{+0.053}$ & $438_{-148}^{+561}$  & $0.99_{-0.16}^{+0.01}$ & 0.8 & \citetalias{Bromberg2013,Berger2014} \T \\ 
GRB~061006  &  0.26  &  0.4377   &  $3.82_{-0.63}^{+0.73}$ & $909_{-191}^{+260}$ & -- & 0.24 & \citetalias{Berger2014} \T \\ 
GRB~061201 & 0.77 & 0.111 & $1.68_{-0.029}^{+0.029}$ & $970_{-209}^{+298}$  & $0.92_{-0.08}^{+0.05}$ & 0.14 & \citetalias{Bromberg2013,Berger2014}\T \\
GRB~070724A  &  0.27  &  0.457   &  $0.016_{-0.003}^{+0.003}$ & $119_{-7.30}^{+7.30}$ & $0.37^{+0.26}_{-0.17}$ & 2.5 & \citetalias{Bromberg2013,Berger2014}\T \\ 
GRB~070809 & 0.44 & 0.2187 & $1.04_{-0.16}^{+0.16}$ & $464_{-223}^{+223}$  & $0.09_{-0.05}^{+0.13}$ & <0.1 & \citetalias{Bromberg2013,Berger2014} \T \\
GRB~071227  &   1.30  &  0.384 & $0.591_{-0.025}^{+0.025}$ & $875_{-287}^{+790}$ & $0.71_{-0.59}^{+0.15}$ & 0.6 & \citetalias{Bromberg2013,Berger2014} \T \\  
GRB~080123  &  0.27  &  0.495  & $3.20_{-1.47}^{+6.59}$ & $2228_{-1308}^{+12723}$ & -- & -- \T \\ 
GRB~080905A  &  0.86  &  0.122  & $0.66_{-0.10}^{+0.10}$ & $658_{-123}^{+293}$ & $0.88_{-0.11}^{+0.07}$ & --  & \citetalias{Bromberg2013}\T \\ 
GRB~100206A  &  0.09  &  0.408 & $0.047_{-0.06}^{+0.06}$ & $708_{-69}^{+0.69}$ & $0.99_{-0.01}^{+0.01}$  &30  & \citetalias{Bromberg2013,Berger2014}\T \\ 
GRB~100625A  &  0.13  &  0.452 & $0.75_{-0.03}^{+0.03}$ & $706_{-116}^{+0.116}$ & $0.97_{-0.03}^{+0.02}$ &0.3  & \citetalias{Bromberg2013,Berger2014} \T \\ 
GRB~130603B$^{\dagger}$  &  0.16  &  0.356 & $1.96_{-0.10}^{+0.10}$  & $823_{-71}^{+83}$ & $0.86_{-0.26}^{+0.26}$ & 1.7 &  \citetalias{Berger2014}\T \\ 
GRB~140903A$^{\dagger}$  &   0.22  &  0.351 & $0.044
_{-0.003}^{+0.003}$ & $60_{-22}^{+22}$  & $0.78_{-0.27}^{+0.08}$ & $1.0\pm0.3$ & \citetalias{host_140903}\T \\ 
GRB~141212A$^{\dagger}$  &  0.19  &  0.596 & $0.068_{-0.011}^{+0.011}$ & $151_{-14}^{+14}$  & $0.78_{-0.07}^{+0.07}$ & --&\T \\ 
GRB~150101B$^{\dagger}$  &  0.02  &  0.093 & $0.0022_{-0.0003}^{+0.0003}$ & $34_{-23}^{+23}$  & $0.86_{-0.09}^{+0.08}$ & $\leq 0.4$&  \citetalias{host_150101b}\T \\ 
GRB~150120A$^{\dagger}$  &  0.8 & 0.46  & $0.19_{-0.04}^{+0.04}$ & $190_{-73}^{+220}$  & $0.33_{-0.10}^{+0.20}$ & -- &\T \\ 
GRB~150423A$^{\dagger}$  &  1.14  &  0.22 &  $0.0075_{-0.001}^{+0.001}$ & $146_{-43}^{+43}$  & $0.83_{-0.10}^{+0.08}$ &-- &  \T \\ 

GRB~160624A$^{\dagger}$  &  0.13  &  0.483 & $0.40_{-0.15}^{+0.14}$ & $1247_{-531}^{+531}$  & $0.84_{-0.07}^{+0.07}$ & -- &\T \\ 
GRB~160821B$^{\dagger}$  &  0.41  &  0.16  & $0.12_{-0.02}^{+0.02}$ & $97.4_{-22}^{+22}$   & $0.67_{-0.10}^{+0.10}$ & --  & \T \\ 
GRB~170428A$^{\dagger}$  &  0.14  &  0.454 & $1.86_{-0.98}^{+0.32}$ & $1428_{-313}^{+346}$  &$0.85_{-0.10}^{+0.09}$ & -- &\T \\ 
GRB~170817A$^{\dagger}$  &  0.50  &  0.00968  &  $4.7e-5_{-0.7e-5}^{+0.7e-5}$ & $65.6_{-14.1}^{+35.3}$  & $0.44_{-0.13}^{+0.15}$& 4e-3 & \citetalias{host_170817A} \T \\ 
\\
\hline
\\
& & & & Group 2 SGRBs & & \\
\\
\hline
\\
GRB~050813  & 0.35  &  0.72  &  $0.15_{-0.08}^{+0.25}$ & $361_{-224}^{+1221}$   & $0.57_{-0.24}^{+0.36}$  &  --  & \citetalias{Bromberg2013,Berger2014}\T \\ 
\\
GRB~060121 & 0.28 & 4.6 & $180_{-12}^{+12}$ & $767_{-67}^{+84}$   & $0.17_{-0.15}^{+0.14}$& --  & \citetalias{Bromberg2013,Berger2014}\T \\  \T
GRB~060801 & 0.33 & 1.131 & $180_{-12}^{+12}$ & $1321_{-439}^{+1379}$    & $0.95_{-0.05}^{+0.03}$ &  6.1  & \citetalias{Bromberg2013,Berger2014} \T \\
GRB~061217 & 0.19 & 0.827 & $4.23_{-0.72}^{+0.72}$ & $731_{-287}^{+895}$ & $0.98_{-0.23}^{+0.01}$ & 2.5 & \citetalias{Bromberg2013,Berger2014}\T \\ 
GRB~070429B &  0.17  &  0.904  &  $0.475_{-0.071}^{+0.071}$ &  $229_{-76}^{+859}$    & $0.32_{-0.15}^{+0.26}$ &  1.1 & \citetalias{Bromberg2013,Berger2014} \T \\ 
GRB~070714B &   0.65  &  0.923  &  $6.4_{-1.1}^{+1.1}$ & $1060_{-215}^{+285}$   & -- & 0.44  & \citetalias{Bromberg2013,Berger2014}\T \\ 
GRB~070729  &  0.56  &  0.8  & $1.13_{-0.44}^{+0.44}$ & $666_{-261}^{+675}$    & $0.89_{-0.57}^{+0.06}$& <0.15  & \citetalias{Bromberg2013,Berger2014}\T \\ 
GRB~090426 &  0.33   &  2.609  &  $8.4_{-1.9}^{+1.9}$ & $1065_{-299}^{+599}$  & $0.10_{-0.06}^{+0.15}$ &  $4.3_{-2.0}^{+2.0}$  & \citetalias{Bromberg2013,Dichiara_2021,Berger2014}\T \\ 
GRB~090510  &  0.51   & 0.903  & $54.6_{-2.1}^{+2.1}$ & $7955_{-343}^{+343}$      & $0.97_{-0.29}^{+0.01}$ & 0.3  & \citetalias{Bromberg2013,Berger2014}\T \\ 
GRB~100117A &   0.27  &  0.915   & $7.8_{-1}^{+1}$  & $547_{-84}^{+84}$     &  $0.97_{-0.03}^{+0.01}$ & <0.2 & \citetalias{Bromberg2013,Berger2014}\T \\   
GRB~101219A  &  0.30   & 0.718   & $6.51_{-0.36}^{+0.36}$   & $1014_{-96}^{+110}$    & $0.94_{-0.06}^{+0.03}$ & --  & \citetalias{Bromberg2013,Berger2014}\T \\ 
GRB~111117A & 0.18   &  2.211   & $8.9_{-3.4}^{+3.4}$   &  $1350_{-450}^{+450}$   & $0.36_{-0.05}^{+0.03}$ &  $17.4_{-6.6}^{+9.4}$  & \citetalias{Bromberg2013,Dichiara_2021,Berger2014}\T \\ 
GRB~120804A &  0.33  &   1.3  & $6.57_{-0.47}^{+0.47}$   & $283_{-41}^{+62}$   & $0.36_{-0.19}^{+0.11}$ &  $40_{-28}^{+33}$ & \citetalias{Dichiara_2021,Berger2014}\T \\ 
GRB~131004A$^{\dagger}$  &  0.90  &  0.71   & $0.69_{-0.03}^{+0.03}$   & $202_{-51}^{+51}$   & $0.24_{-0.07}^{+0.07}$ & -- & \T \\ 
GRB~140622A$^{\dagger}$ &   0.07  & 0.959 & $0.10_{-0.02}^{+0.02}$ & $86.2_{-15.7}^{+15.7}$   & $0.89_{-0.27}^{+0.27}$ & -- \T \\ 
GRB~150424A$^{\dagger}$  &  0.14  &  1.0 & $52.3_{-1.9}^{+1.9}$ &  $1835_{-94}^{+99}$  & $0.59_{-0.22}^{+0.19}$ &-- &\T \\ 
GRB~160410A &  0.58  &   1.717  & $93_{-18}^{+18}$ & $3853_{-973}^{+1423}$   & $0.59_{-0.22}^{+0.19}$ & -- &  \citetalias{Dichiara_2021}\T \\
GRB~200826A$^{\dagger}$  &  0.54  &   0.7486  & $7.09_{-0.28}^{+0.28}$ & $210_{-6.4}^{+6.8}$   & $0.59_{-0.70}^{+0.01}$ & >1.44 \T & \citetalias{zhang_2021} \\
GRB~201221D$^{*}$  &  0.06  &   1.045  & $2.76_{-0.21}^{+0.21}$ & $226_{-35.8}^{+31.8}$   & $0.95_{-0.09}^{+0.09}$ & $2.92_{1.43}^{1.43}$ & This work\T \\
\\
 \hline
 \\
\multicolumn{8}{l}{$^a$ $T_{\rm 90,i}$, $z$, $E_{\rm \gamma, iso}$ and $E_\text{\rm p,i}$ values are taken from \citet{M2019} except for GRB~200826A and \thisgrb}\\
\multicolumn{8}{l}{$^b$ $T_{\rm 90,i}$ = $T_{\rm 90}/(1+z)$}\\
\multicolumn{8}{l}{$^\dagger$ Value of $f_{nc}$ is estimated in the present work}\\
\multicolumn{8}{l}{ $^c$ References for $f_{nc}$ and SFR; \citetalias{Bromberg2013} -- \citet{Bromberg2013}, \citetalias{Berger2014} -- \citet{Berger2014}, \citetalias{Dichiara_2021} -- \citet{Dichiara_2021}, \citetalias{host_140903} -- \citet{host_140903}, \citetalias{host_150101b} -- \citet{host_150101b},    }\\
\multicolumn{8}{l}{\citetalias{host_170817A} -- \citet{host_170817A},\citetalias{zhang_2021} -- \citet{zhang_2021}}\\
\\
\end{longtable}

\bsp	
\label{lastpage}
\end{document}